%% file: CorrectedMain.tex
\documentclass[11pt,a4paper]{article}

\usepackage{graphicx}
\usepackage{bm}
\usepackage{slashed}
\usepackage{braket}
\usepackage[dvipsnames]{xcolor}
\usepackage{jcappub}
\usepackage{amsmath, dsfont}
\usepackage{amsfonts}
\usepackage{amssymb}
\usepackage{setspace}
\usepackage{hypcap}
\usepackage{soul}
\setstcolor{blue}
\newcommand{\lgr}{\mathcal{L}}

\mathcode`*=\string"8000
\begingroup
\catcode`*=\active
\xdef*{\noexpand\textup{\string*}}
\endgroup

\newcommand{\tlambda}{\tilde{\lambda}}

\newcommand{\edit}[1]{{#1}}

\newcommand{\Planck}{\textit{Planck}}

\newcommand{\lcdm}{$\Lambda$CDM}

\newcommand{\beq}{\begin{equation}}
\newcommand{\eeq}{\end{equation}}
\renewcommand{\arraystretch}{1.6}

\def\ee{\end{equation}}
\def\bea{\begin{eqnarray}}
\def\eea{\end{eqnarray}}
\def\bse{\begin{subequations}}

\begin{document}


\title{{Scaling solutions as Early Dark Energy resolutions to the Hubble tension
}}
\author[a]{Edmund J.\ Copeland}
\emailAdd{ed.copeland@nottingham.ac.uk}
\affiliation[a]{School of Physics and Astronomy, University of Nottingham,\\
Nottingham NG7 2RD, United Kingdom}
\affiliation[b]{Institute for Particle Physics Phenomenology, Department of Physics, Durham University,\\ Durham DH1 3LE, United Kingdom}
\author[a]{Adam Moss}
\emailAdd{adam.moss@nottingham.ac.uk}

\author[a,b]{Sergio Sevillano Mu\~{n}oz}
\emailAdd{sergio.sevillano-munoz@durham.ac.uk}

\author[a]{Jade M. M. White}
\emailAdd{jade.white@nottingham.ac.uk}

\abstract{
    
    A wide class of scalar field models including Quintessence and K-essence have the attractive property of tracker regimes, where the energy density stored in the field evolves so as to mimic that of the dominant background component. During this evolution, for a brief period of time, there is an increase in the energy density of the field as it spirals in towards its attractor solution. We show that when the peak of this energy density occurs around the epoch of equality, we can address a key requirement of early dark energy (EDE), postulated as a solution to the Hubble tension. In particular we 
    demonstrate how this can occur in a wide class of Quintessence, axion and K-essence models, before 
    showing that the Quintessence models suffer in that they generally lead to sound speeds incompatible with the requirements of EDE, whereas the K-essence and axion models can do a better job of fitting the data.
}

\maketitle
\newpage
\section{Introduction}\label{intro}
Measuring the expansion rate of our Universe ($H_0$), is without doubt one of the most challenging tasks facing cosmologists \cite{Weinberg:2013agg}. An accurate determination of $H_0$ is essential in obtaining the values of most of our cosmological parameters. It is therefore of little surprise that when two complementary approaches to determine its value fail to agree to within of order five sigma, then there is a concerted effort to understand the differences, and possibly reconcile them \cite{Bernal:2016gxb}. This disagreement between the Hubble parameter determined from direct observations of Cepheids and Type Ia Supernova ($H_0 = 73.04 \pm 1.04~{\rm km}~ {\rm s}^{-1}~{\rm Mpc}^{-1}$) \cite{Riess:2006fw,Riess:2021jrx} and those inferred from measurements of anisotropies in the cosmic microwave background radiation (CMB), assuming a standard six parameter $\Lambda$CDM model, ($H_0 = 67.44 \pm 0.58~{\rm km}~ {\rm s}^{-1}~{\rm Mpc}^{-1}$) \cite{Planck:2018vyg,Efstathiou:2019mdh,Efstathiou:2020wxn} have led to what is known as the Hubble tension. 
Attempts to explain the differences in terms of possible systematics between the two experiments have to date not resolved the discrepancy { (although see \cite{Freedman:2023jcz} for a recent discussion of the systematics associated with the direct determination of the Hubble parameter)}, hence has prompted a renewed interest in the possibility that the tension is present because of new physics that will not have been properly accounted for in the literature to date. The basic idea is that the Planck indirect measurement has been made without the presence of the new physics being included. For recent reviews of possible theoretical models resolving the tension see \cite{Kamionkowski:2022pkx} and \cite{Knox:2019rjx}. The majority of the most promising solutions have the property of increasing the expansion rate of the scale factor just prior to recombination. The go-to resolution for this is by using a dynamical scalar field whose energy density increases rapidly and briefly during this period, thereby allowing the Hubble parameter to increase to that consistent with the Supernova observations -- an effect that has become known as Early Dark Energy (EDE) \cite{Knox:2019rjx}. In particular, it appears that a successful scenario requires that this increase occurring around matter-radiation equality amounts to a brief insertion of around 8\% of the energy density, which then dissipates faster than matter in order that it does not leave any lasting imprints in the CMB. 

Two natural candidates for this behaviour that have to date received relatively little attention come from what are known as Quintessence and K-essence models, scalar field models with dynamics driven by either their potential energy (Quintessence) \cite{Zlatev:1998tr} or non-canonical kinetic energy (K-essence) \cite{Armendariz-Picon:1999hyi,Armendariz-Picon:2000nqq,Chiba:1999ka}. For a review of such dynamical fields see 
\cite{Copeland:2006wr,Bahamonde:2017ize}. 
In this paper, we argue that such models can in fact in principle provide a solution to the Hubble tension. They possess a key property that is known as an attractor solution, where the energy density in the scalar field can mimic that of the background dominating energy density (whether it be radiation or matter) \cite{Copeland:1997et}. In particular, due to the spiral stability of this solution, the field experiences a relative increase in its energy density for a short period while approaching this attractor, sufficient to account for the required EDE impulse. Moreover, as we will demonstrate, a wide variety of models will also feel attracted to this solution at early times, producing the desired EDE peak at equality while decaying at late times.

The layout of the paper is as follows. In section~\ref{section:attractor sols}, as a toy model, we introduce a scalar field $\phi$ with an exponential potential of constant slope $\lambda$, evolving in a $\Lambda$CDM cosmology and show analytically and numerically how it can accommodate a period of EDE. In section~\ref{subsec:fang} we extend this to a more general potential, with a time-dependent slope, allowing us to derive the conditions required for EDE to emerge in general Quintessence models. { In section~\ref{sec:axion} we show how this can work as well for the case of an axion}. In section~\ref{section:k-essence} we turn our attention to the case of K-essence. Considering a particular toy model, where { $K(X) = \frac{X^n}{M^{4n-4}}$, where $X\equiv\frac{1}{2}\dot{\phi}^2$ (where $\dot{\phi}\equiv \frac{d\phi}{dt}$), and $M$ is a mass scale}, (motivated partly by the fact the case $n=2$ leads to a sound speed $c_s^2 = \frac{1}{3}$ in the fluctuations of the K-essence field, a result consistent with the findings in \cite{Moss:2021obd}), we show analytically in a manner similar to the Quintessence case how these too lead to EDE. In section~\ref{Sec:MCMC} we describe the numerical solutions obtained and provide an MCMC likelihood analysis showing the significance of the results. In particular, we show the problems faced by Quintessence models, and the fact that the { K-essence case with $n=3/2$ provides a good fit} to the data. Finally, we conclude in section~\ref{Sec:conclusion}. 


\section{Attractor solutions in Quintessence}\label{section:attractor sols}

In this section, we develop the argument that scalar field evolution in the presence of a background fluid can experience scaling solutions where the energy density of the scalar field aims to become a fixed fraction of that of the dominating background fluid. In following that trajectory, there is a short period of time when the energy density stored in the scalar field itself, increases briefly as it readjusts { to its new scaling regime}. It is this increase that can provide the input required to address the Hubble tension, and in what follows we first of all show the principle of it. 

We begin by introducing the equations of motion, for a system containing a canonical scalar field $\phi$ with potential $V(\phi)$ and two barotropic fluids with energy density $\rho_\gamma$ (radiation) and $\rho_m$, (matter, both baryonic and non-baryonic) with equations of state $\gamma_r=4/3$ and $\gamma_m=1$ respectively, defined in terms of their pressure ($p$) and energy density ($\rho$) by $p= (\gamma-1) \rho$. For completeness we also include a cosmological constant $\rho_{cc} = \frac{\Lambda}{\kappa^2}$ (with an associated equation of state $\gamma_{cc}=0$) to provide the late time dark energy of the universe, although in the analytic analysis below we will drop this term as it is completely sub-dominant around matter-radiation equality, when the effect we are seeking to explain occurs. However, we keep the full equations in the numerical solutions we compare to in section~\ref{Sec:MCMC}.\\

The Friedmann equation is given by: 
\begin{equation}\label{Friedmann}
    H^2=\frac{\kappa^2}{3}\left(\rho_r +\rho_m + \rho_{cc}+\frac{\dot{\phi}^2}{2}+V(\phi)\right),
\end{equation}
where $\kappa=\sqrt{8\pi G}$, $H \equiv \dot{a}/a$ is the Hubble parameter with $a(t)$ the scale factor and $\dot{a} \equiv \frac{da}{dt}$. 
The dynamics and stability of the system will depend on the specific choice for the potential $V(\phi)$. A natural choice is an exponential potential\footnote{Notice that the cosmological constant can be either treated as an extra fluid or absorbed into the potential as a constant term. In the former case, the system is described by the formalism shown in section~\ref{subsec:exponential};  otherwise, we would have to consider a time-dependent $\tlambda(N)$, using the method introduced in section~\ref{subsec:fang}.}, $V(\phi)=V_0 \exp{(-\lambda \kappa \phi)}$, 
with slope parameter $\lambda =$ const, since it presents scaling behavior at late times, as well as the intermediate regime of increased energy density we are searching for.  

\subsection{Exponential potential with a constant slope parameter $\lambda$} \label{subsec:exponential}
The fluid and scalar field equations of motion are
\begin{align}
    \dot{\rho}_r=&-3H\gamma_r\rho_r,\nonumber\\
    \dot{\rho}_m=&-3H\gamma_m\rho_m,\\
    \dot{\rho}_{cc}=&-3H\gamma_{cc}\rho_{cc},\qquad\quad\nonumber\\
    \ddot{\phi}+3H\dot{\phi}& + V,_\phi(\phi)=0.\nonumber
\end{align}
where $V,_\phi(\phi) \equiv \frac{dV}{d\phi}$. Following the prescription introduced in \cite{Copeland:1997et} we convert these equations to first-order ones by introducing, the dimensionless density parameters
\begin{align}\label{parameters}
     x=&\frac{\kappa \dot{\phi}}{\sqrt{6}H}, & y=&\frac{\kappa \sqrt{V}}{\sqrt{3}H}, & z=&\frac{\kappa \sqrt{\rho_r}}{\sqrt{3}H}, 
     & l=&\frac{\kappa \sqrt{\rho_{cc}}}{\sqrt{3}H},
\end{align}\\
which from the Friedmann constraint Eq.~(\ref{Friedmann}) gives the dimensionless energy density in matter via 
\begin{equation}
\label{FriedmannCons}
  \Omega_m \equiv  \frac{\kappa^2 \rho_m}{3H^2} =1-(x^2+y^2+z^2+l^2),
\end{equation}
whilst for completion, we have the important quantity, the dimensionless energy density in $\phi$,
\begin{equation}\label{omega-phi}
    \Omega_{\phi}=\frac{\kappa^2 \rho_\phi}{3H^2} = x^2 + y^2.
\end{equation}
Differentiating the parameters ($x,y,z,l$) with respect to the number of e-folds ($N=\log{a}$), leads to the following closed system (using $\gamma_r=4/3,\gamma_m=1,\gamma_{cc}=0$):
\begin{eqnarray} 
 x'&=& \sqrt{\frac{3}{2}}\lambda y^2 - 
    \frac{x}{2} (3 - 3 x^2 + 3 y^2-z^2-3l^2),\label{x-eqn}\\
     y'&=&-\sqrt{\frac{3}{2}}\lambda x y  + \frac{y}{2}(3 + 3x^2 - 3 y^2 + z^2+3l^2),\label{y-eqn}\\
     z'&=&-\frac{z}{2}(1-3 x^2 + 3 y^2-z^2-3l^2),\label{z-eqn}\\
      l'&=&\frac{l}{2}(3 + 3x^2 - 3 y^2 + z^2+3l^2),\label{l-eqn}
\end{eqnarray}
where $x' \equiv \frac{dx}{dN}$ and we have already substituted the exponential potential with a constant slope parameter $\lambda$,
\begin{equation} \label{V-const-lambda}
    V(\phi)=V_0\exp{(-\kappa \lambda \phi)}.
\end{equation}
To reiterate, here and in  section~\ref{section:k-essence} we drop $\rho_{cc}$ from Eqs.~(\ref{x-eqn}-\ref{l-eqn}) since we are focusing on the effects of the scalar field around matter-radiation equality, where $l^2\ll1$. Although this set of equations allows us to see the evolution of each energy parameter, it proves convenient to introduce the effective equation of state of the background radiation and matter fields, defined via
\begin{equation}
    \gamma_{\rm eff}=1+\frac{p_\gamma +p_m}{\rho_\gamma+ \rho_m}=1+\frac{1}{3}\left(\frac{z^2}{1-x^2-y^2}\right).\label{tau-defn}
\end{equation}
We see that $\gamma_{\rm eff}$ is a particularly useful parameter to use because it only varies between $1 \leq \gamma_{\rm eff} \leq 4/3$, compared to $z$ which varies between 0 and 1. With this in mind, we can replace $z$ in terms of $\gamma_{\rm eff}$ and the system of equations (\ref{x-eqn}) - (\ref{z-eqn}) become
\begin{eqnarray} 
 x'&=& \sqrt{\frac{3}{2}}\lambda y^2 +
    \frac{3x}{2} (-2+2x^2 + \gamma_{\rm eff}(1- x^2 -  y^2)),\label{x-eqn-tau}\\
     y'&=&-\sqrt{\frac{3}{2}}\lambda x y  + \frac{3y}{2}(2x^2 + \gamma_{\rm eff}(1- x^2 -  y^2)),\label{y-eqn-tau}\\
     \gamma_{\rm eff}'&=&(\gamma_{\rm eff}-1) (3 \gamma_{\rm eff}-4).\label{tau-eqn}
\end{eqnarray}
  \begin{figure}
    \centering
    \includegraphics[scale=1.0]{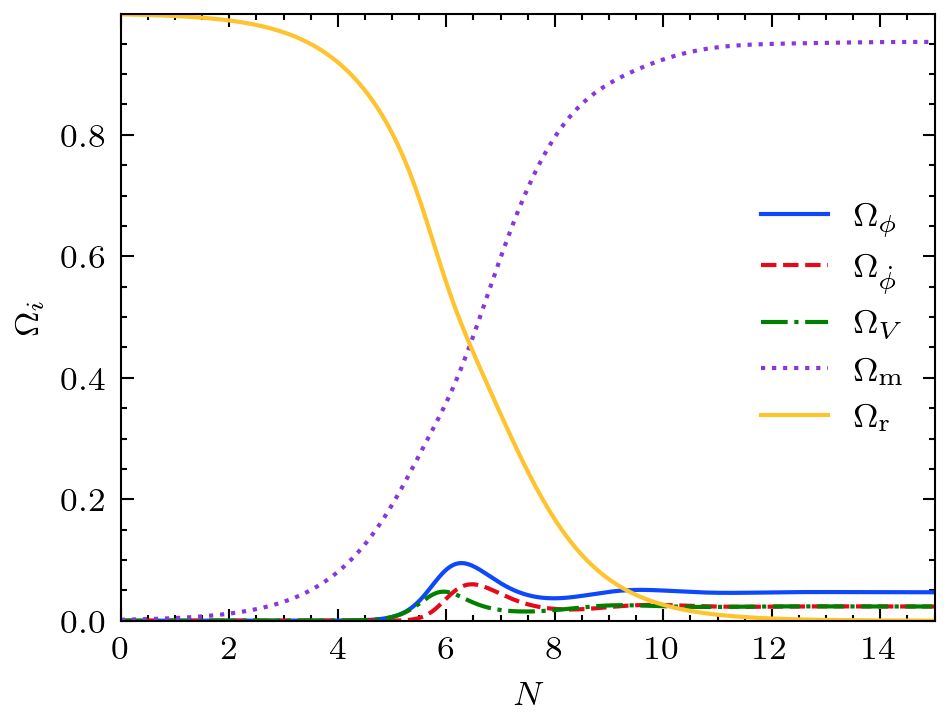}
    \caption{\footnotesize Evolution of a scalar field with an exponential potential Eq.~(\ref{V-const-lambda}) in a background containing matter and radiation barotropic fluids. We can see that during its evolution to the scaling solution fixed point, the field has a local peak in its energy density. The solid yellow line corresponds to $\Omega_r$, the dotted purple line to $\Omega_m$, the solid blue line to $\Omega_\phi$, where as the red dashed and green dashed-dotted lines correspond to the kinetic and potential energy contributions to $\Omega_\phi$ respectively.}  
    \label{fig:ExampleExponential}
\end{figure}
We begin the analysis by noting that throughout both its early and late evolution the scalar field needs to be subdominant, having to satisfy an upper bound at matter domination of $\Omega_\phi < 0.02$ \cite{Planck:2018vyg} (for an example see Figure \ref{fig:ExampleExponential}). From Eq.~(\ref{omega-phi}),  we can therefore neglect terms cubic in $x$ and $y$, implying that Eqs.~(\ref{x-eqn-tau})-(\ref{tau-eqn}) become 
\begin{eqnarray} 
x'&\approx&\left(\frac{3}{2}\gamma_{\rm eff} -3\right)x +\sqrt{\frac{3}{2}}\lambda y^2 ,\label{eq:x_1fluid_exp}\\
     y'&\approx&\frac{3}{2}\gamma_{\rm eff} y-\sqrt{\frac{3}{2}}\lambda x y,\label{eq:y_1fluid_exp}\\
     \gamma_{\rm eff}'&\approx&(\gamma_{\rm eff} -1)\left(3\gamma_{\rm eff} -4 \right).\label{eq:tau_1fluid_exp}
\end{eqnarray}
Note, the nice feature that $\gamma_{\rm eff}$ has fixed points for both matter and radiation domination ($\gamma_{\rm eff}=1$ and $\gamma_{\rm eff}=4/3$, respectively). It is not difficult to show that the fixed point (assuming $\gamma_{\rm eff}$ constant) is given by
\begin{align}\label{eq:fixedpoint}
    x_{\rm sc}=&\sqrt{\frac{3}{2}}\frac{\gamma_{\rm eff}}{\lambda}, & y_{\rm sc}=& \left(\frac{3}{2}\frac{\gamma_{\rm eff}(2-\gamma_{\rm eff})}{\lambda^2} \right)^{1/2}, & \Omega_\phi^{\rm sc} =&\frac{3\gamma_{\rm eff}}{\lambda^2}, & \gamma_\phi =&\gamma_{\rm eff},
\end{align}
  corresponding to the scaling solutions found in ~\cite{Copeland:1997et} (for $\gamma_{\rm eff}=1$ and $\gamma_{\rm eff}=4/3$).  Therefore, depending on the background fluid that is dominating, as long as $\lambda^2 > 3 \gamma_{\rm eff}$, there is a spiral stable attractor solution where $\phi$ evolves so that its energy density tracks that of the dominating background fluid, ruled by $\gamma_{\rm eff}$, behaving as radiation in the early universe, and evolving into matter like behaviour in the matter dominated regime. This is well known~\cite{Copeland:1997et,Barreiro:1999zs}, but there is an interesting element that appears to have been overlooked and could be relevant in addressing the Hubble tension. As shown in Figure \ref{fig:orbit_exp}, due to the spiraling nature of the fixed point, the scalar field will experience oscillations around the attractor in its trajectory. Thus, as these oscillations are damped, the first will lead to a peak in the energy density, which if placed just before matter-radiation equality could alleviate the observed tension. We turn our attention now to determine analytically the properties of the peak, its location in time, and its magnitude in height. 
  
As we can see in Figure \ref{fig:ExampleExponential}, the early-time scalar field energy density is dominated by the potential term ($y$) up to the peak. Moreover, given that to address the Hubble tension the peak must take place before or at matter-radiation equality, the scalar field will be evolving then in a radiation-dominated universe ($\gamma_{\rm eff} \sim 4/3$). Since $x \ll 1$ and $x \ll \lambda$, it follows that before the peak has been reached, Eqs.~(\ref{eq:x_1fluid_exp})-(\ref{eq:y_1fluid_exp}) become 
\begin{eqnarray} 
 x'&\approx&-x+\sqrt{\frac{3}{2}}\lambda y^2 ,\label{eq:x_1fluid_early_exp}\\
     y'&\approx&2 y,\label{eq:y_1fluid_early_exp}
\end{eqnarray}
\begin{figure}
    \centering
    \includegraphics[scale=0.92]{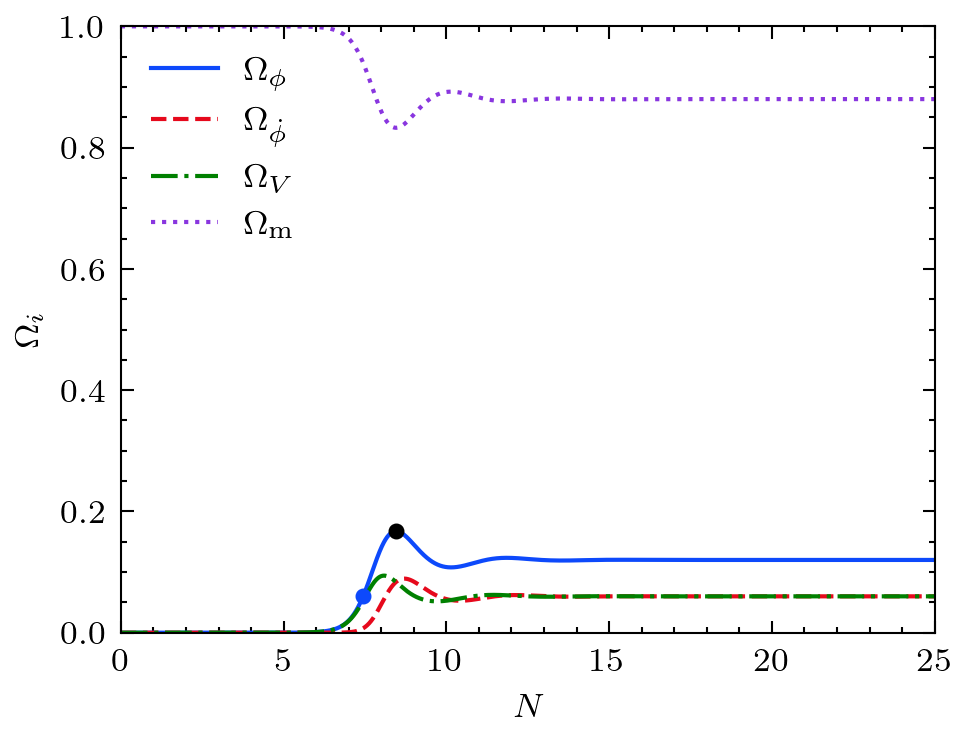}
    \includegraphics[scale=0.92]{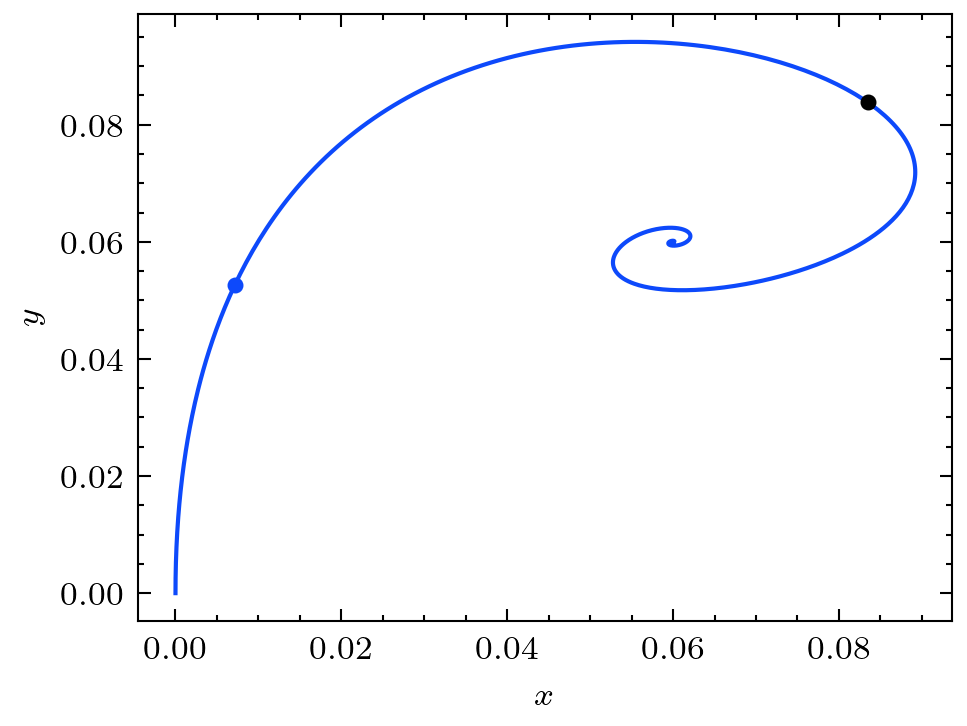}
    \caption{\footnotesize{Evolution of a scalar field with a matter-like fluid ($\gamma_{\rm eff}=1$). The evolution of each of the different densities versus $N$ is represented in the left panel, using the same colour code as in Fig.~\ref{fig:ExampleExponential} while in the right panel we show the corresponding ($x,y$) phase space. Note how the peak (black dot) in $\Omega_\phi$ corresponds to the orbits of the scalar field around the fixed point.}}
    \label{fig:orbit_exp}
\end{figure} 
yielding early-time solutions
\begin{eqnarray} 
  x_{early}(N)&\approx&{(x_i-a_i)e^{-\Delta N_i}+a_i e^{4\Delta N_i}},\label{eq:x_1fluid_exp_early}\\
     y_{early}(N)&\approx&y_i e^{2 \Delta N_i},\label{eq:y_1fluid_exp_early}
\end{eqnarray}
where $\Delta N_i=N-N_i$,  $N_i$ is the initial time, $x_i$ and $y_i$ are the respective initial values and
\begin{equation}
    a_i=\sqrt{\frac{3}{2}}\frac{\lambda y_i^2 e^{4N_i}}{5}.
\end{equation}
These solutions are valid as long as we can drop the higher order terms in Eqs.~(\ref{eq:x_1fluid_exp})-(\ref{eq:y_1fluid_exp}), which takes us close to when the peak in $\Omega_{\phi}$ takes place, a time we call $N_1$. To be more specific we can estimate this time as being the moment $y_{early}(N)$ first passes its final fixed point value Eq.~\eqref{eq:fixedpoint}, which implies (assuming the energy density is equally split between $x$ and $y$),
\begin{equation}
    y_{early}(N_1)\approx \sqrt{\Omega_\phi^{(\rm sc)}/2}.\label{N1definition}
\end{equation}
Using Eqs.~\eqref{eq:fixedpoint} and \eqref{eq:y_1fluid_exp_early}, we obtain the following estimate of the time of the peak
\begin{equation}
    N_{1}\approx N_i +\frac{1}{2}\log{\left(\frac{\sqrt{3\gamma_{\rm eff}}}{y_i \lambda \sqrt{2}}\right)}.\label{N1estimate1}
\end{equation}
Notice that we have reintroduced a generic $\gamma_{\rm eff}$ for $\Omega_\phi^{(\rm sc)}$. This is because the effective equation of state of the universe at the time of the peak ($\gamma_{\rm eff}$) will have an impact on the peak height. The change of $\gamma_{\rm eff}$ would also introduce corrections in the early evolution from Eqs.~(\ref{eq:x_1fluid_exp_early}) and (\ref{eq:y_1fluid_exp_early}), but these are negligible when the time of the peak is around matter-radiation equality.

The evolution of the scalar field from this point to the peak is non-trivial, but it can be approached in two related ways. The first is to perturb the evolution equations about the solutions Eqs.~(\ref{eq:x_1fluid_exp_early}) and (\ref{eq:y_1fluid_exp_early}), evaluated at time $N_1$ to linear order, and solve for the linear fluctuations. The second, related approach is to recognise that we are effectively perturbing around the fixed point solutions Eq.~(\ref{eq:fixedpoint}) at $N_1$ and so we are interested in the stability of these solutions about the fixed point. This stability analysis (assuming $\gamma_{\rm eff}$ is constant) introduces eigenvalues $E_\pm$ with $x(N)$ and $y(N)$ given by  ($N\geq N_1$)
\begin{eqnarray}
     x(N) &=& x_{early}(N_1) + A_x \exp (E_+ \Delta N_1) + B_x \exp(E_- \Delta N_1), \label{x-pertn1}\\
     y(N) &=& y_{early}(N_1) + A_y \exp (E_+ \Delta N_1) + B_y \exp(E_- \Delta N_1),\label{y-pertn1}
 \end{eqnarray} 
 where $\Delta N_1 = N-N_1$,   $A_x,\,B_x,\,A_y$ and $B_y$ are constants with the eigenvalues given by 
\begin{equation}
    E_\pm= -\frac{3(2-\gamma_{\rm eff})}{4}\left(1\pm\sqrt{1-\frac{8\gamma_{\rm eff}}{2-\gamma_{\rm eff}}}\right),\label{eig-value}
\end{equation}
showing that it is a stable spiral. Thus, as the field evolves from its initial position ($\Omega_\phi^{0}\sim 0$) (when $x\ll 1$ and $y\ll 1$) and starts to grow, it feels the fixed point attractor and begins to evolve towards it. In doing so it experiences a damped oscillatory behaviour which is exactly the one that produces the desired peak, as seen in Figure \ref{fig:orbit_exp}.

\begin{figure}
    \centering
    \includegraphics[scale=1.0]{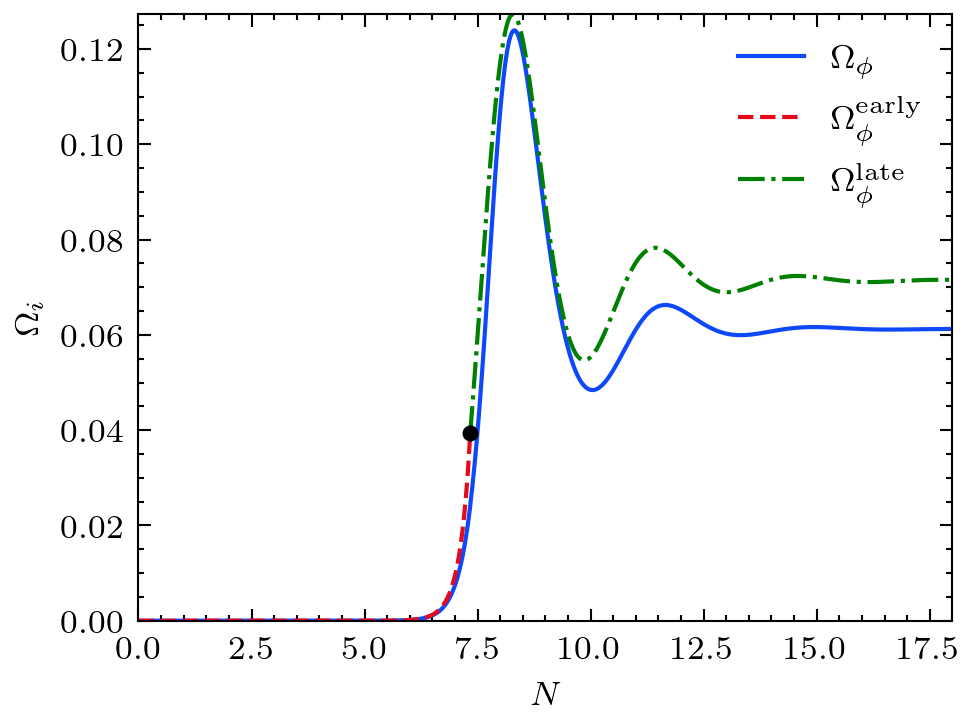}
    \caption{\footnotesize{Comparison of the analytic (red and green lines) and full numerical solution (blue line) for the quintessence potential of Eq.~(\ref{V-const-lambda}) showing the first peak in $\Omega_{\phi}$. The black dot represents the time $N_1$ where the early evolution (dashed red) matches onto the perturbation around the effective fixed point assuming $\gamma_{\rm eff}=7/6$ (dot-dashed green). The analytic solution accurately predicts the first peak of $\Omega_{\phi}$. Beyond the peak, differences emerge as the full numerical system follows the global attractor given by matter domination with} $\gamma_{\rm eff}=1$.}
    \label{fig:predictionQuint1}
\end{figure}

\begin{figure}
    \centering
    \includegraphics[scale=1.0]{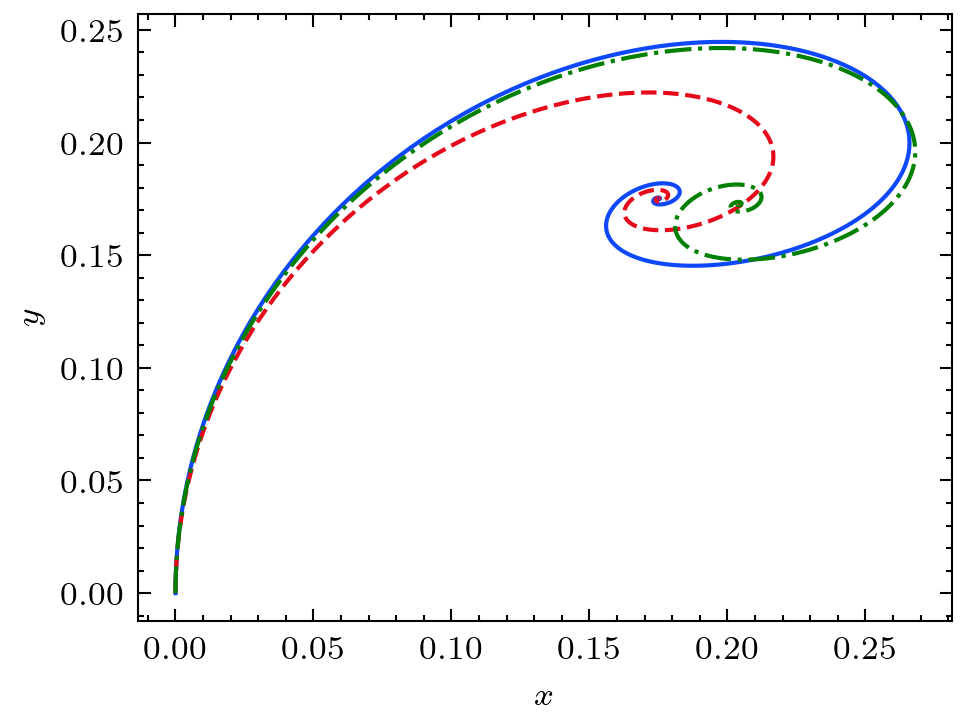}
    \caption{\footnotesize{The $x-y$ orbits of a scalar field around its corresponding fixed point for differing background cosmologies. The blue line represents a scalar field with two background fluids, in which the peak takes place at radiation-matter equality -- the true evolution. The dashed red line corresponds to a scalar field evolving in a matter-dominated background, and the dot-dashed green line to a scalar field in a $\gamma_{\rm eff}=7/6$ background (corresponding to the equation of state of the universe at matter-radiation equality). We can see that the peak in $\Omega_\phi$ (as seen for example in the right pane of Fig.~\ref{fig:orbit_exp}) for the full solution (blue) is well approximated by the effective equation of state of the universe, $\gamma_{\rm eff}=7/6$, before it finally falls down to the late time attractor, given by matter domination.}}
    \label{fig:varying tau}
\end{figure}

A comparison between the analytical approximations derived in this section and the full numerical solution can be seen in the plot highlighting the peak in $\Omega_\phi$ in Figure \ref{fig:predictionQuint1}. They reproduce the time for the peak, $N_1$, and from that time, perturbing around the fixed point evolves through the peak. Notice that the approximation does not match the late-time evolution of the numerical solution. This is because we are perturbing around the fixed point for the value of $\gamma_{\rm eff}$ at the time of the peak, $N_1$. After that, the system will evolve as dictated by the effective attractors at the time. In this way, the field will scale with the dominating background fluid, as we can see in Figure \ref{fig:varying tau}. 

So far, in discussing the nature of the peak in $\Omega_\phi$ we have concentrated on the case of an exponential potential with constant slope parameter $\lambda$, Eq.~(\ref{V-const-lambda}). We have seen how it is possible to understand both the value of the peak height and when it occurs through the fact that starting with $\Omega_\phi \ll 1$, the system wants to head towards its natural scaling solution. However, there is a potential problem, and that is we know the scalar field needs to rapidly become subdominant soon after the peak has been reached, and this is not a natural feature of these models where it is clear from Eq.~(\ref{eq:fixedpoint}) that the system wants to reach a non-negligble and constant  $\Omega_\phi = 3/\lambda^2$. In particular, satisfying the constraint that in the matter-dominated era $\Omega_\phi \lesssim 0.02$ \cite{Planck:2018vyg} implies that $\lambda \gtrsim 12$, which is very constraining. 

Ideally, we would like to have a peak in the energy density driven by the type of mechanism we have described in this section but followed by a decaying $\Omega_\phi$. In section~\ref{subsec:fang}, we show how to achieve this, by considering a model in which the slope parameter of the potential becomes field-dependent, and tends rapidly to large values as the field evolves.

\subsection{Exponential potential with a time dependent slope parameter $\tilde{\lambda}(\phi)$}\label{subsec:fang}
We want to turn our attention to more general Quintessence potentials $V(\phi)$, where the slope parameter, or more accurately the quantity $\frac{V_{,\phi}(\phi)}{V(\phi)}$ can be time dependent. Our aim is to establish the conditions under which such potentials can provide the necessary input of scalar field energy density around equality, whilst satisfying the requirements of EDE. We will follow the basic approach of section~\ref{subsec:exponential}, where the existence of scaling solutions for fixed $\gamma_{\rm eff}$ will play an important role.

With the definitions from Eqs.\eqref{parameters}, we obtain
\begin{eqnarray} 
 x'&=& \sqrt{\frac{3}{2}}\tilde{\lambda} y^2 - 
    \frac{x}{2} (3 - 3 x^2 + 3 y^2-z^2),\label{x-eqn-Vgen}\\
     y'&=&-\sqrt{\frac{3}{2}}\tilde{\lambda} x y  + \frac{y}{2}(3 + 3x^2 - 3 y^2 + z^2),\label{y-eqn-Vgen}\\
     z'&=&-\frac{z}{2}(1-3 x^2 + 3 y^2-z^2),\label{z-eqn-Vgen}\\
    \tilde{\lambda}' &=& -\sqrt{6}\tilde{\lambda}^2(\Gamma-1)x,\label{lambda-eqn-Vgen}
\end{eqnarray}
where, additionally we now have
\begin{align}
 \Gamma=& \frac{V,_{\phi\phi}(\phi)\;V(\phi)}{V,_{\phi}(\phi)^2}, & \tilde{\lambda}=&-\frac{V,_\phi(\phi)}{\kappa V(\phi)}.\label{eq:lambdat and gamma}
\end{align}
Notice that this system is equivalent to the exponential potential case (Eqs.~(\ref{x-eqn}-\ref{y-eqn})) but with a time-varying $\lambda$. An important point to make here is that the nature of the approach to a scaling regime does not rely on having $\tilde{\lambda} =$ constant. In fact, there are many cases where the slope is not a constant, and yet we approach the type of scaling regimes described in Section~\ref{subsec:exponential} - for detailed examples see \cite{Bahamonde:2017ize}. In terms of the $\gamma_{\rm eff}$ parameter defined in Eq.~(\ref{tau-defn}), the system of equations can be written as 
\begin{eqnarray} 
 x'&=& \sqrt{\frac{3}{2}}\tilde{\lambda} y^2 +
    \frac{3x}{2} (-2+2x^2 + \gamma_{\rm eff}(1- x^2 -  y^2)),\label{x-eqn-Vgen-tau}\\
     y'&=&-\sqrt{\frac{3}{2}}\tilde{\lambda} x y  + \frac{3y}{2}(2x^2 + \gamma_{\rm eff}(1- x^2 -  y^2)),\label{y-eqn-Vgen-tau}\\
     \gamma_{\rm eff}'&=&(\gamma_{\rm eff}-1) (3 \gamma_{\rm eff}-4),\label{tau-eqn-Vgen}\\
      \tilde{\lambda}' &=& -\sqrt{6}\tilde{\lambda}^2(\Gamma-1)x,\label{lambda-eqn-Vgen}
\end{eqnarray}
As a particular example, we consider the model of Fang et.~al. \cite{Fang:2008fw}, as it is a nice extension of the exponential potential model, allowing us to understand the evolution of the system in an analogous manner to that case, whilst seeing how it satisfies the requirements of EDE through the time-dependent slope. The model is defined through its potential $V(\phi)$ by 
\begin{align}\label{Gamma-e}
    V(\phi)=V_0 e^{\tilde{\alpha} \phi(\phi+ \beta)/2},& &\tilde{\lambda}=-\tilde{\alpha}(2\phi +\beta)/2,& &\Gamma=1+\frac{\tilde{\alpha}}{\tilde{\lambda}^2},
\end{align}
where $\tilde{\alpha}$ and $\beta$ are constants, and the time-dependent effective slope $\tilde{\lambda}$ satisfies
\begin{equation}\label{lambda-eqn}
 \tilde{\lambda}' = -\sqrt{6}\tilde{\alpha}x.
 \end{equation}
 Applying the same sub-dominant scalar field energy simplifications as in the last section (i.e., considering that $x^2\ll1$ and  $y^2\ll1$) we obtain the following equations describing the system
 \begin{eqnarray} 
 x'&\approx&\left(\frac{3}{2}\gamma_{\rm eff} -3\right)x +\sqrt{\frac{3}{2}}\tlambda y^2,\label{eq:x_1fluid_Fang}\\
     y'&\approx&\frac{3}{2}\gamma_{\rm eff} y,\label{eq:y_1fluid_Fang}\\
     \gamma_{\rm eff}'&=&(\gamma_{\rm eff}-1)(3\gamma_{\rm eff}-4),\label{eq:tau_eqn_Fang}\\
     \tilde{\lambda}' &=& -\sqrt{6}\tilde{\alpha} x,\label{eq:L_1fluid_Fang}
\end{eqnarray}
which for a fixed $\gamma_{\rm eff}$ and $\tilde{\alpha}<0$ has the following stable fixed point
\begin{align}
    x_{\rm sc}=&\; 0, & y_{\rm sc}=&\; 0, & \tilde{\lambda}_{\rm sc} \to &\; \infty.
\end{align}

\begin{figure}
     \centering
     \includegraphics[scale=1.0]{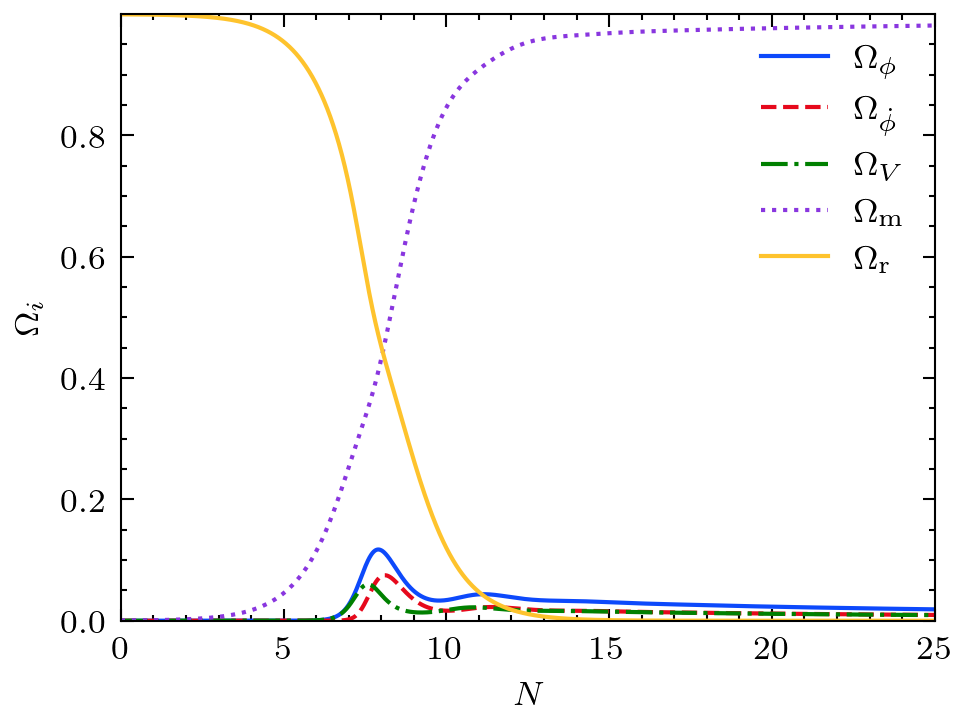}
     \caption{\footnotesize
     Evolution of a scalar field with $V(\phi)=V_0 e^{\tilde{\alpha} \phi(\phi+ \beta)/2}$, $\tilde{\alpha}<0$, in a background containing both matter and radiation barotropic fluids. Although the fixed point in this case is at $x=y=0$, the system develops a peak during its trajectory around the period of radiation-matter equality.}
     \label{fig:ExampleFang}
 \end{figure}

\begin{figure}
    \centering
    \includegraphics[scale=0.9]{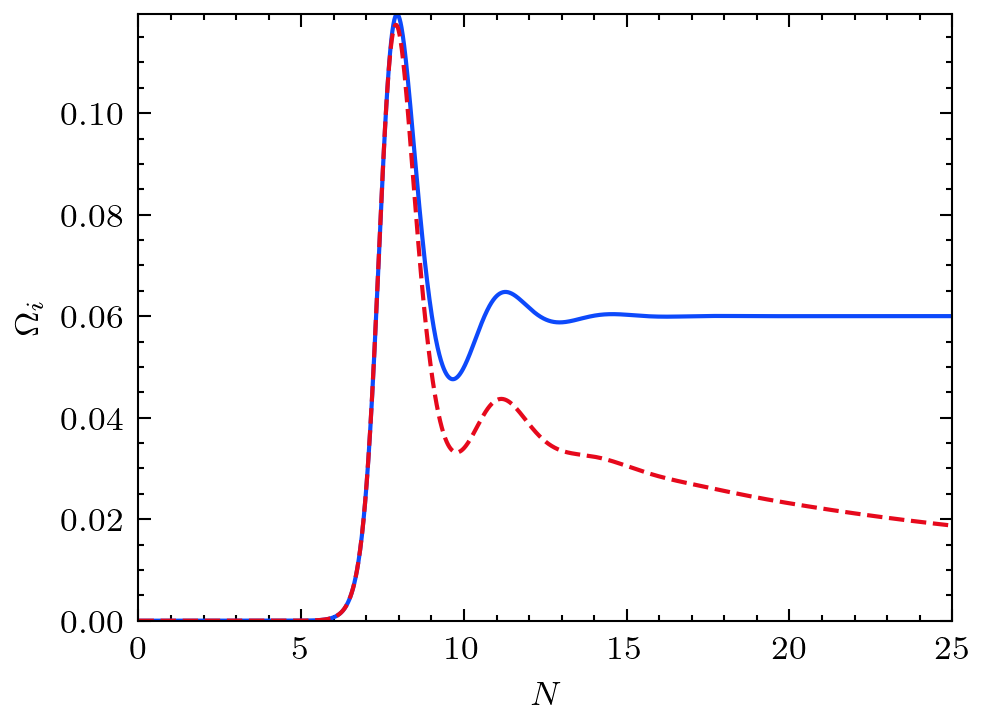}
    \includegraphics[scale=0.9]{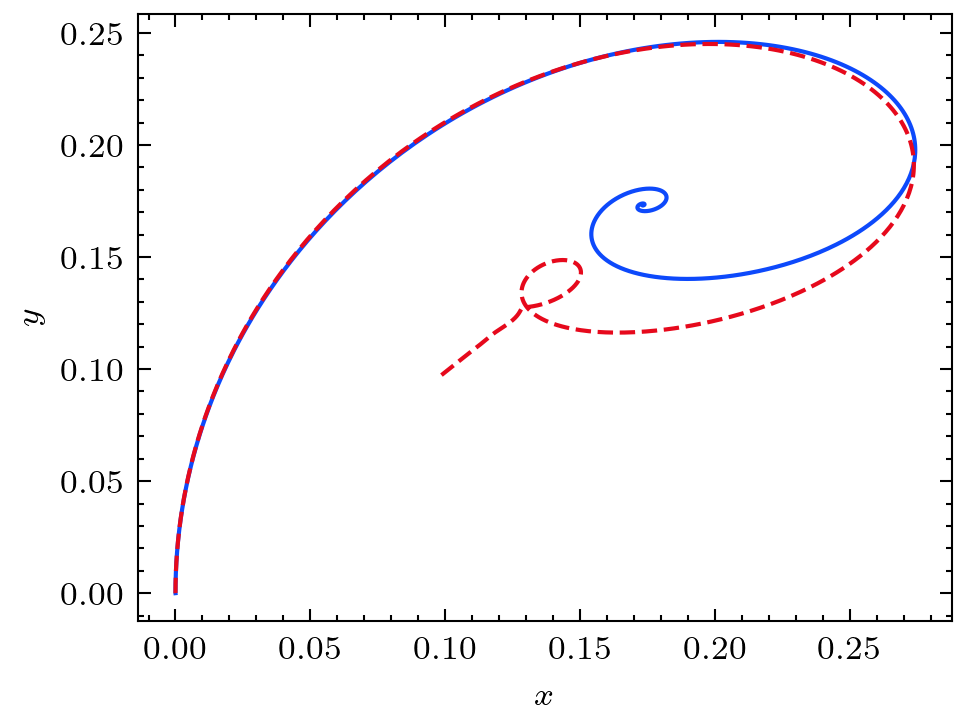}
    \caption{\footnotesize{Evolution of a scalar field with a matter-like fluid ($\gamma_{\rm eff}=1$). The evolution of each of the different densities versus $N$ is represented in the left panel, while the right panel shows the $x-y$ phase space during this evolution. We can see that for the model with evolving $\tilde{\lambda}$ given by potential Eq.~(\ref{Gamma-e}) (dashed red), there is always a pure exponential case with constant $\tilde{\lambda}$ that matches the trajectory up to the peak (blue).}}
    \label{fig:FangvsExp}
\end{figure}

Given that the initial conditions are $x_i \ll 1$ and $y_i \ll 1$, we might naively expect no evolution in $x$ and $y$, hence in $\Omega_\phi$, implying no peak forming, since they are already close to their fixed points. However, that does not take into account the evolution of $\tilde{\lambda}$ whose initial value $\tilde{\lambda}_i$ is a long way from its global fixed point, which, as we see from Figure \ref{fig:ExampleFang}, can be crucial. Moreover, given that the evolution equation for $\tlambda$ is  $\tilde{\lambda}'\propto x$, it implies that $\tlambda\approx$ const, until close to the peak.  
The only difference between the exponential case of section~\ref{subsec:exponential} and this more general case is the fact that $\tlambda$ begins to vary slowly as $x$ begins to evolve, leading eventually to a changing effective fixed point once the peak is approached. 
This implies that we expect the same behaviour as for the exponential case of section~\ref{subsec:exponential} up to $N_1$. The early time solutions for $x$ and $y$ are once again given by Eqs.~(\ref{eq:x_1fluid_exp}-\ref{eq:y_1fluid_exp}), and it is only when $x$ begins to grow that we need to take into account the evolution of $\tilde{\lambda}$. This is a really nice feature, it means that we can always find a solution for the exponential case that fits the early evolution (and the peak) of more generic potential models, as long as $\tilde{\lambda}$ is not varying rapidly. As another example we will see this in section~\ref{sec:axion} for the case of the axion. It is worth stressing that this is more general than the specific result for the potential Eq.~(\ref{Gamma-e}), it holds because $\tilde{\lambda}$ remains essentially constant up to the peak, which is a common feature of Quintessence as they will always have $\tilde{\lambda}'\propto x$ (see Eq.~\eqref{lambda-eqn-Vgen}).
 In particular Eq.~(\ref{N1estimate1}) gives $N_1$ where we use $\tilde{\lambda}=\tilde{\lambda}_i$ as a leading approximation, to obtain
\begin{equation}
    N_{1}\approx N_i +\frac{1}{2}\log{\left(\frac{\sqrt{3\gamma_{\rm eff}}}{y_i \tlambda_i \sqrt{2}}\right)}.\label{N1estimatefang}
\end{equation}
We should note here that there is nothing special about matter-radiation equality leading to the peak here. It means that there is a degree of finetuning required in order to make the system evolve towards its peak value. 
However, for the perturbation around the scaling fixed point, we can obtain a better estimate by solving Eq~(\ref{eq:L_1fluid_Fang}) with the early time solution for $x$ namely, Eq.~(\ref{eq:x_1fluid_exp_early}), with initial conditions $\tilde{\lambda}(N_i)= \tilde{\lambda}_i$.
and with $\lambda \to \tilde{\lambda}_i$ in Eq.~(\ref{eq:x_1fluid_exp_early}). When we do this we obtain, 
\begin{equation}
    \tlambda_{early}(N)=\tlambda_i -\sqrt{6}\tilde\alpha \left[\frac{a_i}{4}\left(e^{4\Delta N_i}-1\right)-(1-a_i)\left(e^{-\Delta N_i}-1\right)\right],\label{lambda-evolve}
\end{equation}
where now $\Delta N_i \equiv N - N_i$.
Given we have determined $\tlambda_{early}(N_1)$, in Figure \ref{fig:FangvsExp}, we show how well an exponential potential with constant $\lambda = \tlambda_{early}(N_1)$ solution matches the true evolution up to and beyond the first peak in $\Omega_\phi$. Therefore, we can apply the same bounds on $\tlambda_{early}(N_1)$ as for the exponential potential case, meaning that to address the Hubble tension we need $\tlambda_{early}(N_1)\approx 8$. After the peak takes place, the exponential case approximation 
quickly fails as $\tlambda$ keeps on increasing towards its fixed point at $\tlambda\to\infty$. However, as we know from the stability analysis, the scaling solution for the scalar field implies that its energy density vanishes at late times. 

Based on a combination of analysing the behaviour of the system of equations (\ref{x-eqn-Vgen-tau})-(\ref{lambda-eqn-Vgen}) and the particular case of the Fang et.~al. model \cite{Fang:2008fw}, we can make some general statements about the properties of scalar field potentials that can lead in principle to a successful period of EDE. A scalar field will be a good candidate for the Hubble tension if it satisfies the following: 
\begin{itemize}
    \item[1)] \textbf{Before the peak:} A predictable evolution of $\tlambda(N)$ that ends in a value of $\tlambda(N_1)\approx\edit{8}$, which will produce the required peak \edit{${\Omega_\phi}^{(\rm sc)}\approx 0.1$} due to the orbits around the effective scaling solution. 
    \item[2)] \textbf{At the peak:} A $\tlambda(N)$ that doesn't vary too rapidly around $N_1$, so the field has enough time to orbit around the fixed point. 
    { In particular for 10\% accuracy, we require $7.9< \tlambda(N_1) < 8.1$ over a time interval $\delta N_1$ of roughly half the orbital period to the fixed point. This is related to the imaginary part of Eq.~(\ref{eig-value}) and corresponds to $\delta N_1\approx 4 \pi/(3\sqrt{7})$~e-folds.   }
   
    \item[3)] \textbf{After the peak:} A $\tlambda(N)$ that tends to infinity, so that the effective scaling solution for the system goes back to $x=0$ and $y=0$ at late times, corresponding to a subdominant $\Omega_\phi$. \\
    
\end{itemize}
In the following subsection, we focus on an axion model in which $\tlambda(N)$ varies very rapidly around the peak. We will show that although this makes it difficult to accurately predict the height of the peak, we are still able to use this method to explain the systems early and late-time evolution. 

 \subsection{The axion potential}\label{sec:axion}
\begin{figure}
     \centering
     \includegraphics[scale=1.0]{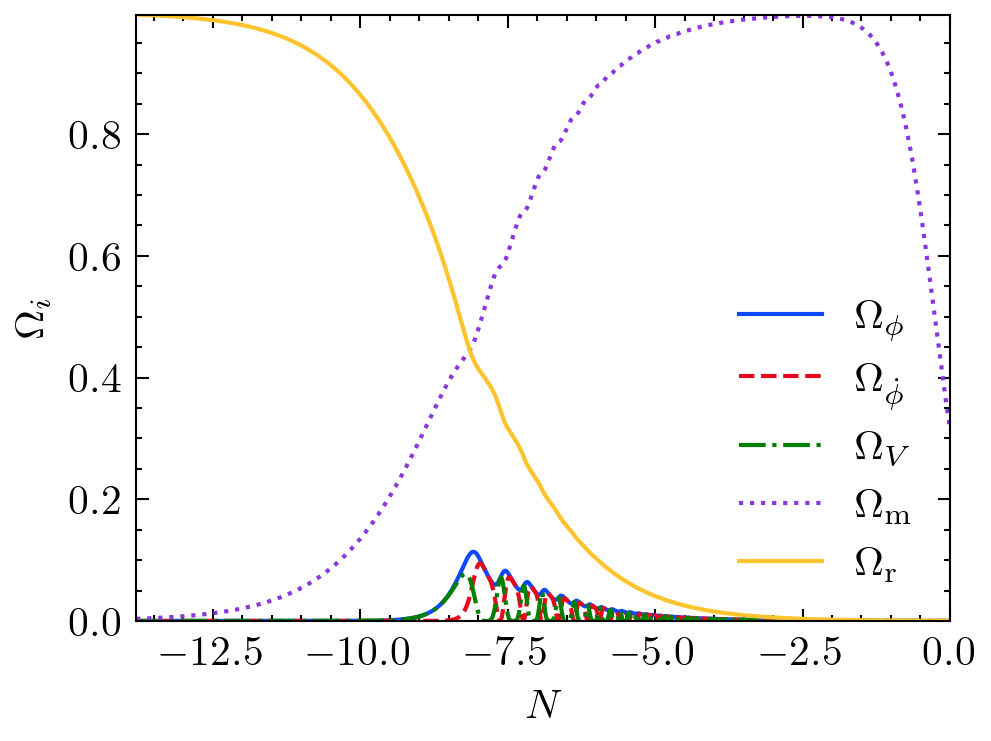}
     \caption{\footnotesize
     Evolution of a scalar field with an axion potential Eq.~(\ref{eq:axion-potential}) with $m=2$.}
     \label{fig:axion}
 \end{figure}

A major drawback with using slowly rolling Quintessence models, as an explanation of EDE is that they lead to a sound speed $c_s^2 =1$ for the field $\phi$, and as we will see in section~\ref{Sec:MCMC}, the data seems to favour $c_s^2 < 1$ around the time of EDE \cite{Knox:2019rjx,Kamionkowski:2022pkx,Moss:2021obd}. Fortunately, as we will now show, the techniques we have developed can be applied to the case of a rapidly oscillating field with $c_s^2 <1$, such as the case of the axion. Consider a generic axion field evolving under the following potential
\begin{equation}\label{eq:axion-potential}
V_{{\rm Ax}}^{(m)}(\phi)=\mu_m^4 \left(1-\cos\left(\frac{\phi}{f_m}\right)\right)^m,
\end{equation}
where $\mu_m$ and $f_m$ are constants. Using the parametric redefinitions from Eq.~\eqref{parameters}, we find the following equations of motion\footnote{For the axion, we find it is numerically more stable to solve for the variable $1/\tilde{\lambda}$ due to the large increase in $\tilde{\lambda}$.}
\begin{eqnarray} 
 x'&=& \sqrt{\frac{3}{2}}\tilde{\lambda} y^2 - 
    \frac{x}{2} (3 - 3 x^2 + 3 y^2-z^2),\label{x-eqn-ax}\\
     y'&=&-\sqrt{\frac{3}{2}}\tilde{\lambda} x y  + \frac{y}{2}(3 + 3x^2 - 3 y^2 + z^2),\label{y-eqn-ax}\\
     z'&=&-\frac{z}{2}(1-3 x^2 + 3 y^2-z^2),\label{z-eqn-ax}\\
    \tilde{\lambda}' &=& \frac{\sqrt{6}x}{2m}\left(\tlambda^2+\frac{m^2}{f_m^2\kappa^2}\right).\label{lambda-eqn-ax}
\end{eqnarray}
which with Eq.~\eqref{eq:lambdat and gamma}, leads to the following expressions for the axion potential
\begin{align}
  \tilde{\lambda}=&-\frac{m}{\kappa f_m}\frac{\sin\left(\frac{\phi}{f_m}\right)}{1-\cos\left(\frac{\phi}{f_m}\right)}, &
   \Gamma=&1-\frac{1}{2m}-\frac{m}{2f_m^2\kappa^2\tlambda^2}.
\end{align}
As we can see in Figure~\ref{fig:axion}, the axion model also produces a peak in its energy density that can alleviate the Hubble tension. Focusing on the equation of motion for $\tlambda(N)$, we can see that our analysis from Section~\ref{section:attractor sols} can be expended to study the evolution of the axion potential. This is because, since $\tlambda'$ depends linearly on $x$, the value for $\tlambda$ will be frozen at early times, when $x\ll1$. Therefore, the system can be approximated by an exponential potential solution at early times, where $\tlambda'\to0$. 

Connecting this system to our previous analysis, the early evolution for the axion can be approximated to that of an exponential potential until the energy density of the scalar field is of the same order as its corresponding scaling solution. For $\tlambda_i<m/f_m\kappa$, we can linearize Eq.~\eqref{lambda-eqn-ax}, allowing us to follow the same steps as in section~\ref{subsec:fang}, and obtain the early evolution for $x,y$ and $\tlambda$. After some algebra (and substituting for $\gamma_{\rm eff}=4/3$), we obtain 
\begin{align}
  x_{early}(N)&\approx{(x_i-a_i)e^{-\Delta N_i}+a_i e^{4\Delta N_i}},\label{eq:x early axion}\\
     y_{early}(N)&\approx y_i e^{2 \Delta N_i},\\
    \tlambda_{early}(N)&\approx\tlambda_i+\frac{m}{f_m \kappa}\tan\left(\frac{3\tlambda_i y_i^2}{40 f_m \kappa}e^{4\Delta N_i}\right),\label{eq:lam early axion}
\end{align}
where $\Delta N_i=N-N_i$. As expected, the only difference in the early evolution of different models is in $\tlambda_{early}(N)$, which will affect the height of the subsequent peak in energy density. Therefore, we can use these equations to calculate the value for $N_1$, which marks the end of the period in which the early-time approximation is valid, and so the time of the peak. This quantity was introduced in Eq.~\eqref{N1estimate1}, and is given by 
\begin{equation}
    N_{1}\approx N_i +\frac{1}{2}\log{\left(\frac{\sqrt{3\gamma_{\rm eff}}}{y_i \tlambda_i \sqrt{2}}\right)},
\end{equation}
where, $\gamma_{\rm eff}$ is the effective equation of state of the universe at the time of the peak. Notice that, as mentioned in section~\ref{subsec:fang}, the fine-tuning in Quintessence EDE models is exactly of the same order as for the axion fields. 
 \begin{figure}
     \centering
     \includegraphics{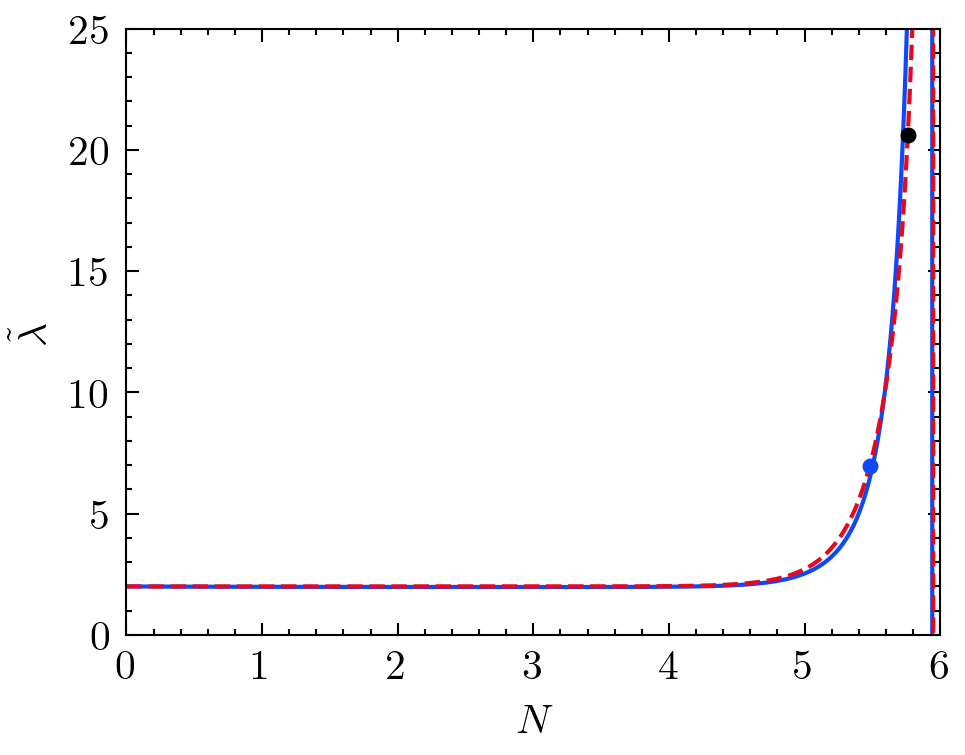}
     \caption{\footnotesize
     Comparison between the numerical evolution $\tlambda(N)$ (blue) against the analytical approximation from Eq.~\eqref{eq:lam early axion} (red dashed) with $\tlambda_i=2$ and $f=0.1$. The black point marks the predicted value for $N_1$ and the subsequent $\tlambda_1$, which should be substituted into an exponential potential solution to mimic the peak for the full system. The blue point represents the correct value for $\tlambda_1$ needed to have a perfect match between the exponential and the axion peak. Although the analytical expression for $\tlambda_{early}(N)$ matches the evolution of $\tlambda(N)$ at early times, we see that the steep slope of the function at $N_1$ leads to a small error on $N_1$ which in turn leads to a large difference in $\tlambda_1$. }
     \label{fig:lambda axion}
 \end{figure}
Although the early evolution for $x,y$ and $\tlambda$ is well described by Eqs.~(\ref{eq:x early axion}-\ref{eq:lam early axion}) (as shown in Figure~\ref{fig:lambda axion}), the rapid evolution of $\tlambda_{early}(N)$ at $N=N_1$ makes it very difficult to make an analytic prediction of the scalar field's energy density parameter at the peak. \edit{This is a perfect example for a violation of the second condition:} even though we have a very good approximation for $N_1$ (and so the time at which the peak takes place), the steep slope of $\tlambda_{early}(N)$ at $N_1$ implies that any uncertainty in the value of $N_1$ will have a significant impact on $\tlambda_{early}(N_1)$ due to its exponential dependence on $N$ (see Eq.~\eqref{eq:lam early axion}). Therefore, given that the peak height is very sensitive to the value of $\tlambda_{early}(N_1)$, it is not possible (without the use of iterative methods to improve the value for $N_1$) to make accurate analytical predictions on the behaviour of the axion field during the peak. Nevertheless, it is still possible to estimate the late-time evolution of the field's total energy once the peak has taken place.  
For this, we just need to consider that once the axion field unfreezes, it will start oscillating around the minimum of its potential. During these oscillations, there are periods in which either $x$ or $y$ dominate the field's energy content, extending the oscillations to the equation of state for the axion field. However, we can average the fluctuations into an effective equation of state that approximates the decay of the scalar field. For this, we must first make the following approximation (valid as long as the field is close to the minimum) to the potential
\begin{equation} \label{min:axionpot}
 V_{{\rm Ax}}^{(m)}(\phi)\approx\mu_m^4 \left(\frac{\phi^2}{2 f_m^2}\right)^m.
\end{equation}
This is necessary as we can now apply the virial theorem to estimate the average ratio between $x$ and $y$ during these oscillations, which, depending on the exponent in the potential ($m$), is given by $\left\langle x^2\right\rangle\approx m\left\langle y^2\right\rangle$. Thus, the effective equation of state for the axion field can be averaged to
\begin{equation}
\left\langle\gamma_\phi\right\rangle=\frac{2\left\langle x^2\right\rangle}{\left\langle x^2\right\rangle+\left\langle y^2\right\rangle}\approx\frac{2m}{m+1}.
\end{equation}
This result agrees with the previous literature~\cite{Poulin:2018dzj,Marsh:2015xka} and shows that in order for the axion field to act as dark energy we need $m\leq2$, such that it decays faster than radiation. 

In summary, we can predict the behavior for a scalar field with a generic potential as long as its associated $\tlambda$ varies slowly. For this, we just need to find the value for $\tlambda$ at the time of the peak, and use the same analysis and equations we used in the exponential case but using the approximated $\tlambda_1$. The underlying idea of finding a $\lambda$ constant case that matches the early evolution of Quintessence fields can be extended to a class of models with non-canonical kinetic energies, known as K-essence models. In the next section, we will show how they too can lead to observationally viable periods of EDE while satisfying the bound on its sound speed, namely $c_s^2<1$. 

\section{K-essence case: $\lgr(X,\phi)=X^n-V(\phi)$}\label{section:k-essence}

So far we have concentrated on the evolution of a canonical scalar field in the early universe, asking how it can address the Hubble tension. To date, relatively little attention has been paid to the role non-canonical fields could play, yet these are known to have some very interesting cosmological properties and arise in a number of particle inspired settings \cite{Armendariz-Picon:1999hyi,Armendariz-Picon:2000nqq,Chiba:1999ka,Malquarti:2003nn,Scherrer:2004au,Das:2006cm,Jorge:2007zz,Bahamonde:2017ize} although there have been questions raised over their ability to act as dark energy \cite{Bonvin:2006vc}. We are going to consider it here as a way of providing successful EDE. With that in mind, we consider a class of simplified such models, with Lagrangian's of the form
\begin{equation} 
    \lgr= \frac{X(\dot{\phi})^n}{M^{4(n-1)}} -V(\phi),\label{non-can-lag}
\end{equation}
where $X(\dot{\phi})\equiv\frac{1}{2}\dot{\phi}^2$, $M$ is a mass scale introduced to keep the action dimensionless and $n$ is a constant. Of course, $n=1$ corresponds to a canonical scalar field. One of the interesting aspects of these models is that they lead to reduced sound speeds of the field $\phi$. In particular, we find for this case \cite{Pourtsidou:2013nha,Skordis:2015yra}
\begin{equation}
    c_s^2=\frac{1}{2n-1},\label{cssq}
\end{equation}
depending on the exponent in the kinetic energy function $X(\dot{\phi})$. A number of the proposals for early dark energy have included resolutions that include reduced sound speeds, for example see \cite{Poulin:2018dzj, Moss:2021obd}. In what follows, we keep $n$ general, allowing us to constrain the full parameter space of $n$.  

For completeness, we begin by once again introducing the equations of motion for our system, which now contains the non-canonical Lagrangian Eq.~(\ref{non-can-lag}), plus the two barotropic fluids introduced in section~\ref{section:attractor sols}, with energy density $\rho_r$ (radiation) and $\rho_m$, (matter, both baryonic and non-baryonic) and equations of state $\gamma_r$ and $\gamma_m$ respectively \cite{Bahamonde:2017ize}.
The Friedmann equation is given by 
\begin{equation}\label{eq:Friedmann-Kessence}
    H^2=\frac{\kappa^2}{3}\left(\rho_r +\rho_m +\frac{2n-1}{2^n M^{4(n-1)}}(\dot{\phi}^2)^n+V(\phi)\right),
\end{equation}
while the fluid and scalar field equations of motion are
\begin{align}
    \dot{\rho}_r=&-3H\gamma_r\rho_r,\nonumber\\
    \dot{\rho}_m=&-3H\gamma_m\rho_m,\\
    \frac{n(2n-1)}{2^{n-1}M^{4(n-1)}}\;\dot{\phi}^{2n-2}\;\ddot{\phi} +&\frac{3Hn}{2^{n-1}M^{4(n-1)}}\;\dot{\phi}^{2n-1}+ V,_\phi(\phi)=0.\nonumber
\end{align}
We may introduce the dimensionless density parameters (note the new definition of $x$ here)
\begin{align}\label{parameters-Kessence}
     x=&\frac{\kappa \sqrt{2n-1}}{ 2^{n/2}M^{4(n-1)}}\frac{\dot{\phi}^n}{\sqrt{3}H}, & y=&\frac{\kappa \sqrt{V}}{\sqrt{3}H}, & z=&\frac{\kappa \sqrt{\rho_r}}{\sqrt{3}H}, 
\end{align}\\
which when substituted into Eq.~(\ref{eq:Friedmann-Kessence}), gives the dimensionless density in matter
\begin{equation}
\label{FriedmannCons}
  \Omega_m \equiv  \frac{\kappa^2 \rho_m}{3H^2} =1-(x^2+y^2+z^2),
\end{equation}
with the dimensionless density in $\phi$,
\begin{equation}\label{omega-phi-K-essence}
    \Omega_{\phi}=\frac{\kappa^2 \rho_\phi}{3H^2} = x^2 + y^2.
\end{equation}
For consistency with the earlier sections we concentrate on the case of the exponential potential $V=V_0 \exp(-\kappa \lambda \phi)$ in Eq~(\ref{V-const-lambda}), such that the evolution equations become 
(with $\gamma_r=4/3,\gamma_m=1$):
\begin{eqnarray}
x'&=&\sqrt{\frac{3}{2}} \frac{M^{4(n-1)}\lambda y^2}{ (2 n-1)^{\frac{1}{2 n}}}\left(\frac{\kappa}{\sqrt{3}H x}\right)^{1-\frac{1}{n}} +\frac{x}{2} \left(\frac{3 x^2-3}{2n-1}-3 y^2+ z^2\right),\label{x-eqn-kess-n}\\
y'&=&-\sqrt{\frac{3}{2}} \frac{M^{4(n-1)}\lambda xy}{ (2 n-1)^{\frac{1}{2 n}}}\left(\frac{\kappa}{\sqrt{3}H x}\right)^{1-\frac{1}{n}} +\frac{y}{2} \left(3+\frac{3 x^2}{2 n-1}-3 y^2+z^2\right),\label{y-eqn-kess-n}\\
z'&=&\frac{z}{2}\left(-1+\frac{3x^2}{2n-1}-3 y^2+z^2\right),\label{z-eqn-kess-n}
\end{eqnarray}
where recall $x' \equiv \frac{dx}{dN}$. Once again working in terms of $\gamma_{\rm eff}$, the effective equation of state of the background fluids defined in Eq.~(\ref{tau-defn}), we have
\begin{eqnarray}
x'&=&\sqrt{\frac{3}{2}} \frac{M^{4(n-1)}\lambda y^2}{ (2 n-1)^{\frac{1}{2 n}}}\left(\frac{\kappa}{\sqrt{3}H x}\right)^{1-\frac{1}{n}}+\frac{3x}{2} \left(\frac{2n( x^2-1)}{2n-1}+\gamma_{\rm eff}(1-x^2- y^2)\right),\\
y'&=&-\sqrt{\frac{3}{2}} \frac{M^{4(n-1)}\lambda xy}{ (2 n-1)^{\frac{1}{2 n}}}\left(\frac{\kappa}{\sqrt{3}H x}\right)^{1-\frac{1}{n}} +\frac{3y}{2} \left(\frac{2n x^2}{(2 n-1)}+\gamma_{\rm eff}(1- x^2-y^2)\right),\\
\gamma_{\rm eff}'&=&(\gamma_{\rm eff}-1)(3\gamma_{\rm eff}-4).
\end{eqnarray}
Note, however, that this system of equations is not obviously closed due to the explicit dependence of the system on $H(N)$. It proves convenient to combine $H(N)$ into a new parameter $\eta$ such that the system manifestly becomes explicitly closed, 
\begin{equation}
    \eta= \frac{ M^{4(n-1)}\lambda}{ (2 n-1)^{\frac{1}{2 n}}}\left(\frac{\kappa}{\sqrt{3}H x}\right)^{1-\frac{1}{n}},
\end{equation}
leading to the following system of equations 
\begin{eqnarray}
x'&=&\sqrt{\frac{3}{2}}\eta y^2+\frac{3x}{2} \left(\frac{2n( x^2-1)}{2n-1}+\gamma_{\rm eff}(1-x^2- y^2)\right),\label{xpgenn}\\
y'&=&-\sqrt{\frac{3}{2}} \eta xy+\frac{3y}{2} \left(\frac{2n x^2}{(2 n-1)}+\gamma_{\rm eff}(1- x^2-y^2)\right),\label{ypgenn}\\
\gamma_{\rm eff}'&=&(\gamma_{\rm eff}-1)(3\gamma_{\rm eff}-4),\label{taupgenn}\\
\eta '&=&\eta  (n-1) \left(\frac{3}{2 n-1}-\sqrt{\frac{3}{2}}\frac{ \eta  y^2}{n x}\right).  \label{etapgenn}  
\end{eqnarray}
We note the close resemblance between these equations and the particular set of Quintessence equations Eqs.~(\ref{eq:x_1fluid_Fang}-\ref{eq:L_1fluid_Fang}), where the only differences are in the form of the evolution of $\eta$ (corresponding to the time-dependent $\tilde{\lambda}$ in Quintessence) and one of the terms in the evolution of $x$. The natural late time evolution of the system Eqs.~(\ref{xpgenn})-(\ref{etapgenn}) for fixed $\gamma_{\rm eff}$ is   
\begin{align}\label{eq:fixedpoint_eta_n}
    x_{\rm sc}& \to 0, & y_{\rm sc}& \to 0, & \eta_{\rm sc}&= \sqrt{\frac{2}{3}}\left(\frac{3n}{2n-1}\right)\frac{x}{y^2} \to \infty.
\end{align}
Therefore, naively, for a system that starts close to the fixed points $x=y=0$ we expect little evolution. However, as we saw in Section \ref{subsec:fang}, even though $x$ and $y$ start close to their late time fixed points, $\eta$ (in that case $\tilde{\lambda}$) starts far from it, and this is what leads to a non-trivial evolution of the field.  
\begin{figure}
    \centering
    \includegraphics[scale=1.0]{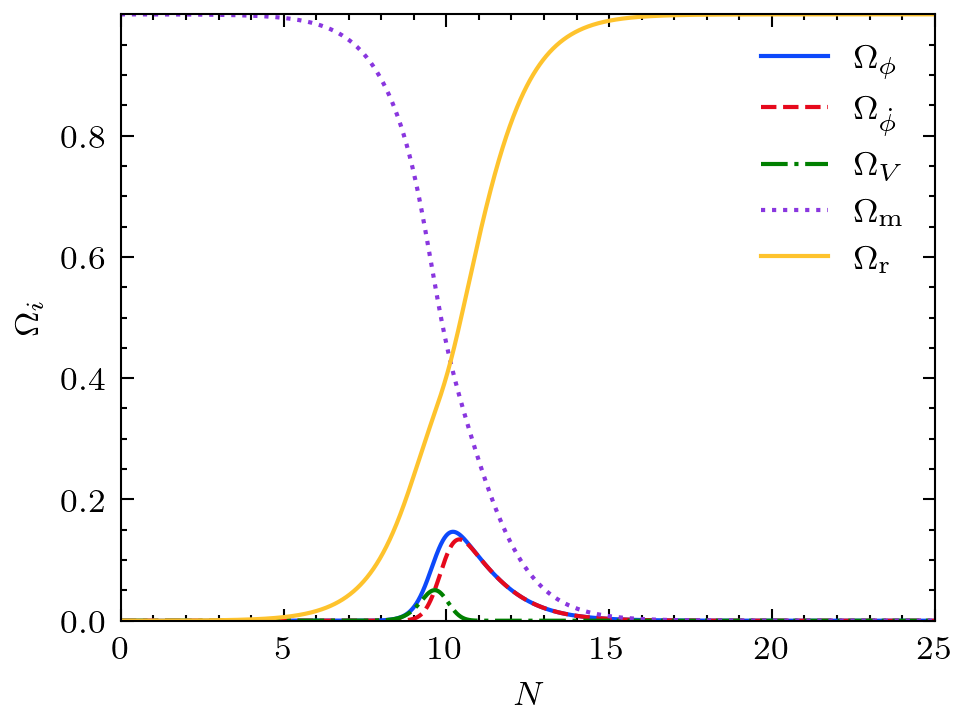}
    \caption{\footnotesize Evolution of a k-essence ($X^2$) scalar field with an exponential potential in a background containing both matter and radiation baryotropic fluids. Although the late time fixed point is at $x=y=0$, the system presents a peak during its trajectory as $\eta$ begins to grow.}
    \label{fig:k-essence-example}
\end{figure}
In fact, the similarity with the set of Quintessence equations Eqs.~(\ref{eq:x_1fluid_Fang}-\ref{eq:L_1fluid_Fang}) doesn't end there. It suggests that we should be able to use a similar approach to that adopted in section~\ref{subsec:fang} for $\tilde{\lambda}$. In this way, we may find an evolution for a system with a constant $\eta$ case that matches the full evolution of K-essence (with varying $\eta$) up to the peak, in a similar manner to Quintessence where the exponential case matched the Fang et al. model in Section \ref{subsec:exponential}. 

For the case where we have a constant $\eta =\eta_c$, equations (\ref{xpgenn})-(\ref{etapgenn}) reduce to
\begin{eqnarray}
x'&=&\sqrt{\frac{3}{2}}\eta_c y^2+\frac{3x}{2} \left(\frac{2n( x^2-1)}{2n-1}+\gamma_{\rm eff}(1-x^2- y^2)\right),\label{xpgenn-const-eta}\\
y'&=&-\sqrt{\frac{3}{2}} \eta_c xy+\frac{3y}{2} \left(\frac{2n x^2}{(2 n-1)}+\gamma_{\rm eff}(1- x^2-y^2)\right),\label{ypgenn-const-eta}\\
\gamma_{\rm eff}'&=&(\gamma_{\rm eff}-1)(3\gamma_{\rm eff}-4),\label{taupgenn-const-eta}
\end{eqnarray}
which has the following spiral stable fixed point (assuming a constant $\gamma_{\rm eff}$ and $n\leq\frac{2\gamma_{\rm eff}}{2\gamma_{\rm eff}-1}$)\footnote{Notice that in a radiation dominated universe ($\gamma_{\rm eff}=4/3$) this fixed point ceases to exist for $n\geq2$, which will be used later to rule out a whole range of K-essence models.}
\begin{align}\label{eta-const-fixedpoints}
    x_{\rm (sc)}=&\sqrt{\frac{3}{2}}\frac{\gamma_{\rm eff}}{\eta_c}, & y_{\rm (sc)}=&\sqrt{\frac{3}{2}}\frac{{ \sqrt{\gamma_{\rm eff}}}}{\eta_c}\sqrt{\frac{2n}{(2n-1)}-\gamma_{\rm eff}}, & \Omega_\phi^{\rm (sc)}=&\frac{3 n \gamma_{\rm eff}}{(2n-1)\eta_c^2}, & \gamma_\phi^{\rm(sc)}=&\gamma_{\rm eff}.
\end{align}
As inferred from the equation of state, this solution corresponds to the scaling fixed point for this class of K-essence models (reducing to Quintessence for $n=1$). To find the corresponding constant value for $\eta_c$ in Eqs.~(\ref{xpgenn-const-eta}-\ref{taupgenn-const-eta}) that perfectly describes the peak of the full system, we just need to calculate the value for $\eta$ close to the peak, at $N=N_1$, which we call $\eta_1$.

\begin{figure}
    \centering
    \includegraphics[scale=1.0]{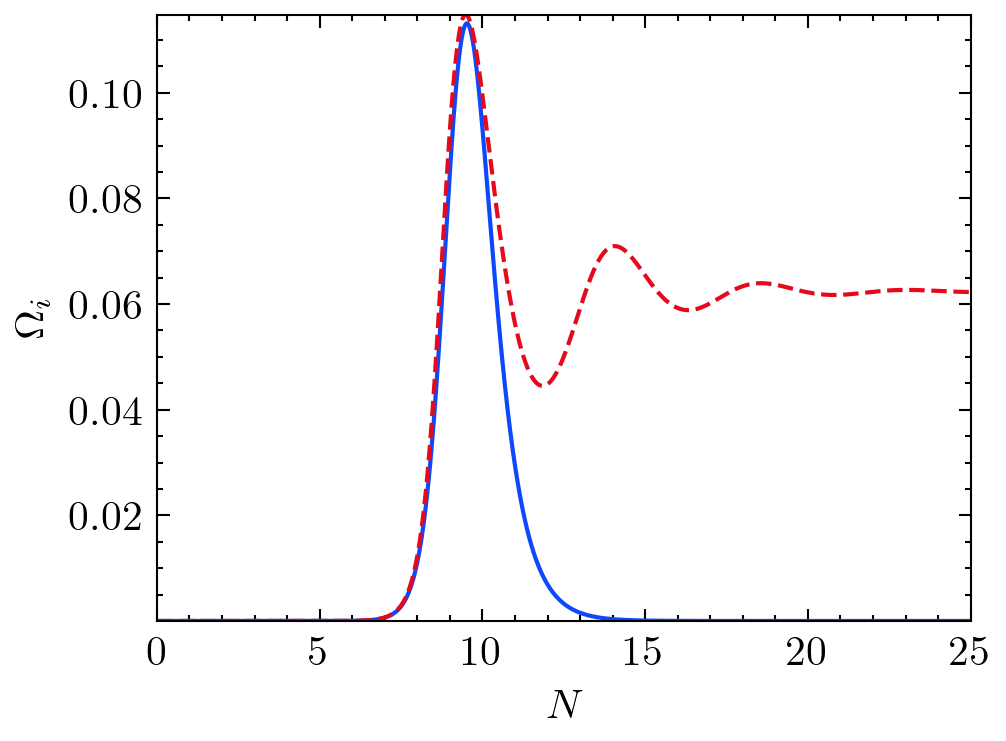}
    \caption{\footnotesize{Comparison between the constant $\eta_c=\eta_1$ case, (Eqs.~(\ref{xpgenn-const-eta}-\ref{taupgenn-const-eta})) (red-dashed line) and the full system (blue line) for K-essence with $n=3/2$ (both systems solved numerically). We can see that the full system tracks the first peak of the scaling solution for a constant $\eta_c$, decaying back to the origin after the peak takes place.}}
    \label{fig:etaconstvsfull}
\end{figure}

To relate the value for $\eta_1$ to its initial value $\eta_i$, we will study the case for a set of initial conditions in which both $x$ and $y$ start very close to the origin, and $\eta_i \sim O(1)$, such that $x\ll 1,\; y<1,\; y < \eta_i$. To address the Hubble tension, we know that the peak must take place at matter-radiation equality or slightly before. Therefore, for this early evolution of the fields, we can assume that the universe is radiation dominated, meaning that $\gamma_{\rm eff}=4/3$. With this, Eqs.~(\ref{xpgenn})-(\ref{etapgenn}) simplify to
\begin{eqnarray} 
 x'&=& \left(2-\frac{3n}{2n-1}\right)x+\sqrt{\frac{3}{2}}\eta y^2,\label{x-eqn-eta-Kess-n}\\
y'&=&2y-\sqrt{\frac{3}{2}} \eta xy,\label{y-eqn-eta-Kess-n}\\
\eta '&=&\eta  (n-1) \left(\frac{3}{2 n-1}-\sqrt{\frac{3}{2}}\frac{ \eta  y^2}{n x}\right)  , \label{eta-eqn-eta-Kess-n}
\end{eqnarray} which can be solved to give (by dropping the $\eta x y$ term in Eq.~(\ref{y-eqn-eta-Kess-n}))
\begin{eqnarray}
x_{early}(N)&=& x_i e^{2 \Delta N_i}\left((1-b_i)e^{-3\Delta N_i}+b_i e^{2 \Delta N_i}\right)^{\frac{n}{2n-1}},\label{x-eqn-early-kess-n}\\
y_{early}(N)&=&y_i e^{2\Delta N_i}\label{y-eqn-early-kess-n}\\
\eta_{early}(N) &=&\eta_i \left((1-b_i)e^{-3\Delta N_i}+b_i e^{2 \Delta N_i}\right)^{-\frac{n-1}{2n-1}},   \label{eta-eqn-early-kess-n}  
\end{eqnarray}
where $\Delta N_i=N-N_i, x_i=x(N_i), y_i=y(N_i), \eta_i=\eta(N_i)$ and
\begin{equation}
b_i=\sqrt{\frac{3}{2}}\frac{(2n-1)}{5n}\frac{y_i^2 \eta_i}{x_i}.
\end{equation}
Eqs.~(\ref{x-eqn-early-kess-n}-\ref{eta-eqn-early-kess-n}) provide the early time evolution of the system. These equations are valid as long as we can ignore the higher order term in Eq.~\eqref{y-eqn-eta-Kess-n}, which becomes important when $y_{early}(N)$ is in the vicinity of the scaling fixed point, at a time we call $N_1$. Therefore, as for Quintessence, we may define this time as when the following equality is satisfied
\begin{equation}
    y_{early}(N_1)\approx\sqrt{\Omega^{\rm(sc)(N_1)}_\phi/2}=\sqrt{\frac{3 n \gamma_{\rm eff}}{2(2n-1)\eta_{early}(N_1)^2}},
\end{equation}
where we can see that the fixed point is time-dependent, given the evolution of $\eta(N)$. Substituting for $y_{early}(N_1)$ and $\eta_{early}(N_1)$ from Eqs.~\eqref{y-eqn-early-kess-n} and \eqref{eta-eqn-early-kess-n}, we find
\begin{equation}\label{N1-eta}
    N_1=N_i +\frac{2n-1}{2n}\log\left(\sqrt{\frac{3n\gamma_{\rm eff}}{2(2n-1)}}\frac{b_i^{\frac{n-1}{2n-1}}}{y_i\eta_i}\right).
\end{equation}
Therefore, the value for $\eta$ at the time of the peak, $\eta_1$, is given by
\begin{equation}
    \eta_1=\eta_i \left(\frac{x_i}{y_i}\right)^{\frac{n-1}{n}}\left(\frac{10}{3}\sqrt{\frac{n}{\gamma_{\rm eff}(2n-1)}}\right)^{\frac{n-1}{n}},
\end{equation}
which we can use for the constant $\eta_c$ scenario in Eqs.~(\ref{xpgenn-const-eta}-\ref{taupgenn-const-eta}) to match the peak of the full system. In Figure~\ref{fig:etaconstvsfull} we can see the specific case for $n=3/2$ and $\gamma_{\rm eff}=7/6$, giving a nearly perfect approximation.
Therefore, to calculate the trajectory for the constant $\eta$ system from $N_1$ to the peak,
we just need to perturb around the fixed point solutions shown in Eq.~(\ref{eta-const-fixedpoints}) for $\eta_c=\eta_1$. Performing the stability analysis leads to \begin{eqnarray}
     x(N) &=& x_{early}(N_1) + C_x \exp (E_+ \Delta N_1) + D_x \exp(E_- \Delta N_1), \label{x-pertn-eta}\\
     y(N) &=& y_{early}(N_1) + C_y \exp (E_+ \Delta N_1) + D_y \exp(E_- \Delta N_1),\label{y-pertn-eta}
 \end{eqnarray} 
 with $C_x,\,D_x,\,C_y$ and $D_y$ being a new set of constants, and for generic $n$, we obtain for the eigenvalues $E_\pm$
 \begin{equation}
    E_\pm=\frac{3}{4}\left(\gamma_{\rm eff}-\frac{2n}{2n-1}\right)\left(-1\pm\sqrt{1-\frac{8\gamma_{\rm eff}}{\frac{2n}{2n-1}-\gamma_{\rm eff}}}\right),
\end{equation}
which are complex with a negative real part for $1\leq\gamma_{\rm eff}<4/3$. 
\begin{figure}
    \centering
    \includegraphics[scale=1.0]{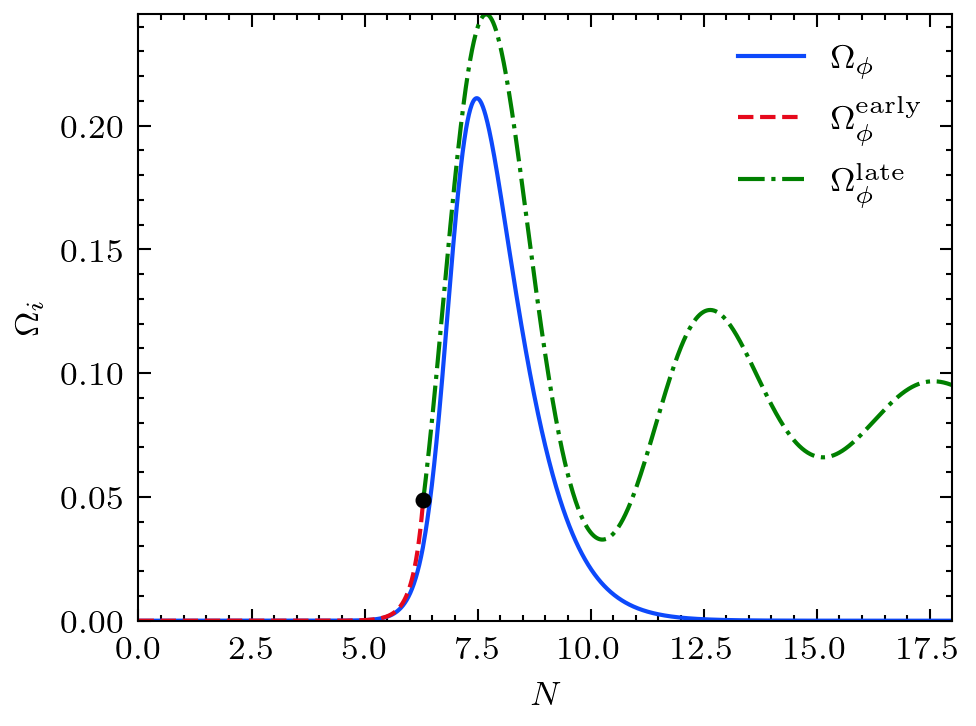}
    \caption{\footnotesize{Comparison of our semi-analytic versus full numerical solution for K-essence with $n=3/2$. The black dot represents $N_1$ where the early analytic evolution (dashed red line) matches the perturbation around the effective fixed point at $\gamma_{\rm eff}=7/6$ and $\eta_c=\eta_1$ (dot-dashed green). We can see that the analytic procedure can describe with a precision of $10\%$ the magnitude of the first peak of the scalar field. After that, both solutions begin to diverge from each other as the full system (blue) follows the global attractor given by $\gamma_{\rm eff}=1$ and $\eta\to\infty$, vanishing faster than radiation.}}
    \label{fig:predictionKess}
\end{figure}

In Figure \ref{fig:predictionKess} we can see a comparison between our semi-analytic prediction and the full numerical calculation. As for Quintessence, our analytic predictions only take us to the peak, since the late time evolution will be determined by the global stability of the system, which we recall is at $x=0,y=0$ (given in \edit{Eq.~\eqref{eq:fixedpoint_eta_n}}). The rate of decay for the scalar field from the peak to this fixed point can be calculated by noticing that in general the kinetic energy term, $x$, dominates the energy density of the scalar field after the peak.
Thus, since the effective equation of state for a K-essence field with a Lagrangian of the form Eq.~(\ref{non-can-lag}) is given by
\begin{equation}
    \gamma_\phi=\frac{2n x^2}{(2n-1)(x^2+y^2)},
\end{equation}
we know that the equation of state of the decaying scalar field ($y\approx 0$) is
\begin{equation}
    \gamma_\phi=\frac{2n}{(2n-1)}.
    \label{gphi}
\end{equation}
Notice that this equation of state is not unique to the exponential potential case, since the potential density parameter $y$ already vanishes by the time $\Omega_\phi$ starts decaying. This result will allow us to place an upper bound on K-essence models in the following section, where we will also compare these models against observations, finding the equilibrium point between a low speed of sound and a large equation of state is preferred by the data. 


\input{results}

\section{Conclusion and discussion}\label{Sec:conclusion}
In this paper we have taken seriously the prospect that the current Hubble tension is a manifestation of new early universe physics. In particular, an evolving scalar field which for a short period of time, around the period of matter radiation equality could briefly enhance the energy density of the universe, and by doing so increase the Hubble parameter beyond the CMB derived value ($H_0 = 67.44 \pm 0.58~{\rm km}~ {\rm s}^{-1}~{\rm Mpc}^{-1}$) \cite{Planck:2018vyg,Efstathiou:2019mdh,Efstathiou:2020wxn} and closer to that determined by the SH0ES team ($H_0 = 73.04 \pm 1.04~{\rm km}~ {\rm s}^{-1}~{\rm Mpc}^{-1}$) \cite{Riess:2006fw,Riess:2021jrx}. In doing so, we of course recognise, that this may be overkill, the resolution may reside in the way the data is analysed and interpreted -- for a recent discussion of the issues concerning direct measurements see \cite{Freedman:2023jcz}). 
The scalar field models that have been proposed to date, have in common a short period just before matter-radiation equality where the energy density increases briefly to around 5-10\% of the overall energy density $\Omega_\phi \sim 0.1$, before reducing again faster than radiation, so as to make sure it does not impact on the matter dominated era that then follows (for a review of models see \cite{Knox:2019rjx,Kamionkowski:2022pkx}).  

The novel element in this work is the recognition that for many scalar field models, evolving in the presence of a background dominating fluid, they will approach a scaling regime, where the energy density stored in the field mimics that of the fluid and becomes a fixed fraction of the energy density. In particular, we ask under what conditions will such a field lead to the required conditions just described around the time of matter-radiation equality. We have addressed this question for a class of exponential potential models (Eq.~(\ref{V-const-lambda}) with a constant slope parameter $\lambda$, Eq.~(\ref{Gamma-e}) for a time-dependent slope $\lambda(\phi)$), Eq.~(\ref{eq:axion-potential}) for an axion field and then for a particular class of K-essence models (Eq.~(\ref{non-can-lag})). In each case we have been able to obtain solutions, where starting from initial conditions corresponding to $x\ll1$ and $y\ll1$ we find that as the system evolves, the field initially remains approximately constant before naturally trying to evolve towards the scaling regime it would associate with matter domination. It is in that evolution, that there is a brief period where the energy density increases allowing the field to act like EDE. We explained analytically the evolution in all four cases and showed how to understand both the evolution of $\Omega_\phi$, and the epoch of domination in terms of the potential parameters and initial conditions of the field. We saw that the standard Quintessence model with constant $\lambda$ could not provide the necessary enhancement in $\Omega_\phi$ whilst also satisfying matter domination constraints on its energy density, but the time-dependent $\lambda$ quintessence models and the axion model could, as their energy density dropped rapidly just after equality. 

A set of conditions were established for such models to act as EDE. Moreover, we were able to demonstrate a nice relationship between Quintessence and the K-essence models, which were also successful in leading to the required EDE. \edit{However, we find that there is nothing special about matter-radiation equality leading to the necessary peak, meaning that there is a degree of finetuning required for these models to relax the Hubble tension, as can be seen in Eq.~\eqref{N1estimatefang}}. The real test of the models though lies in how well they fit all the available data, including the evolution of the associated fluctuations of the field and the formation of structure such as in the CMB. Using an MCMC likelihood fit, we saw that the K-essence model with $n \approx 3/2$ was a much better fit providing an improved $\chi^2$ fit of 14.3 compared to standard $\Lambda$CDM (see Table~\ref{table:parameters}). Although the variable $\lambda$ model of Fang et al \cite{Fang:2008fw}, provided an excellent fit to the background EDE cosmology, it failed eventually because of the fact all Quintessence models have a sound speed $c_s^2 =1$ which the data just does not prefer. 

Of course we are not pretending that the K-essence model we have analysed is a particularly well-motivated particle physics model. However, it does bring home the possibility that interesting dark energy physics can emerge from such non-canonical terms as they can provide ideal realisations of the constraints emerging on EDE solutions from the background field cosmology, i.e. $\gamma_\phi > \frac{4}{3}$ just after matter-radiation equality, and the perturbations of the fields, i.e. $c_s^2 <1$ during that period. It would be fascinating to see whether such models exist in the world of particle physics.
\section*{Note added}
 Just as this paper was being completed, we became aware of \cite{Ramadan:2023ivw} in which the authors also make use of the fact that dynamical systems can lead to enhanced energy densities as they approach their fixed points. In their case, they attempted to use the same field for both EDE and late time dark energy.
\section*{Acknowledgments}

We are grateful to Eoin \'{O} Colg\'{a}in and Vivian Sabla for useful correspondence. The work of EJC and AM was supported by an STFC Consolidated Grant [Grant No. ST/T000732/1], and EJC was also supported by a Leverhulme Research Fellowship [RF-2021 312]. SSM was supported by an STFC studentship [Grant No.\ ST/V506928/1] and 
STFC Consolidated Grant No.~ST/T001011/1, and JMMW by an STFC studentship [Grant No.\ ST/W507702/1]. 

    \end{document}

%% file: results.tex
\section{Comparison with observations}
\label{Sec:MCMC}

The analytic arguments presented in sections \ref{section:attractor sols} and \ref{section:k-essence}, suggest that for there to be viable EDE models associated with the tracking regimes of Quintessence and K-essence, we require initial conditions where $x \ll 1$ and $y \ll 1$, in order to establish that a peak value $\Omega_\phi \sim 0.1$ can occur around matter-radiation equality. We first confirm this by solving the full numerical solutions for all three cases analysed in sections \ref{section:attractor sols} and \ref{section:k-essence}, namely Quintessence with constant slope parameter $\lambda$, with time-dependent $\lambda (\phi)$, (including the case of the axion field) and K-essence with constant slope parameter $\lambda$ and constant $n$. Herein we present a summary of the so-far obtained bounds imposed by observations on each case: 
\begin{itemize}
    \item \textbf{Quintessence with constant $\lambda$:} This case is highly constrained by observations. Although it is possible to obtain a peak height at equality for small enough $\lambda$, we know the scalar field needs to rapidly become subdominant soon after the peak has been reached, and this is not a natural feature of these models. From Eq.~(\ref{eq:fixedpoint}) the system wants to reach a non-negligible and constant  $\Omega_\phi = 3/\lambda^2$.  This was confirmed in full numerical simulations showing these models are inconsistent as EDE models. We note that similar conclusions have been reached previously in the context of both standard~\cite{delaMacorra:2020zqv} and assisted~\cite{Sabla:2021nfy} quintessence to address the Hubble tension. 
    \item \textbf{Slow rolling Quintessence with time-dependent $\tlambda(N):$} These models, along with axion models were discussed in section~\ref{subsec:fang} and section~\ref{sec:axion} respectively, where we confirmed numerically that it was possible to both obtain the correct peak and late time behaviour of $\Omega_\phi$. This is encouraging, however, we know that Quintessence models generally have $c_s^2=1$, and EDE favours lower sound speeds~\cite{Knox:2019rjx,Kamionkowski:2022pkx,Moss:2021obd}.
    \item \textbf{K-essence with $K(X)=X^n$:} The K-essence model we analysed in section \ref{section:k-essence} indicates both analytically and numerically that it is possible to have successful models for EDE peaking around matter-radiation equality for $1 \leq n \leq 2$ and a range of values of the slope parameter $\lambda$. One of the promising features of these models is that, from Eqns.~(\ref{cssq}) and (\ref{gphi}), in the small $x$ and $y$ regimes, $c_s^2$ and the effective equation of state ($\gamma_\phi$) of the field depend on $n$. In particular we see that for $1 \leq n\leq 2$, then $1 \geq c_s^2 \geq \frac{1}{3}$ and $2 \geq \gamma_\phi \geq \frac{4}{3}$. These two conditions bring complementary constraints on the favoured values of $n$. Given the data prefers $c_s^2 <1$, this suggests we require $n>1$, which effectively rules out standard quintessence models (as already discussed), whereas demanding that the energy density in $\phi$ falls off at least as fast as radiation suggests $n \leq 2$. 
\end{itemize}

With this in mind, we now turn our attention to the likelihood of these models acting as EDE. To begin the analysis we first integrate the autonomous set of equations into the public \textsc{Camb} code~\citep{Lewis:1999bs}. This was performed in two steps: First, cosmological parameters from \textsc{Camb} (notably the radiation and dark energy densities today, ($\Omega_{\rm r,\,0}$ and $\Omega_{\rm DE,\,0}$, with $\Omega_{\rm DE,\,0}=\Omega_{\rm cc,\,0}+\Omega_{\phi,\,0}$) were passed to a separate module to solve the autonomous system $(x', y', \gamma_{\rm eff}', \tilde{\lambda}', l')$ for Quintessence with variable $\lambda$, and $(x', y', \gamma_{\rm eff}', \eta', l')$ for K-essence. The system was evolved from $N_i=\ln(a_i)$ with $a_i=10^{-6}$, with initial values $x_i$, $y_i$, $\tilde{\lambda}_i$, $\eta_i$ chosen from wide enough flat priors to encompass the full range of peak redshifts and magnitudes. The initial conditions $\gamma_{\rm eff, i}$ and $l_i$ were found by a bisection search to give the desired $\Omega_{\rm r,\,0}$ and $\Omega_{\rm DE,\,0}$. From this, $\gamma_{\phi}(z)$ and $c_s^{2}$ (which can be calculated analytically for the models we consider) are inputted to \textsc{Camb} to solve for the full cosmological evolution, including perturbations. We note there is a small inconsistency in this approach, as the autonomous equations do not include massive neutrinos in the background evolution. However, we found that when modeling neutrinos as 2 massless species, and 1 massive species with $m_{\nu}=0.06\,{\rm eV}$, there is only a small difference in the resulting $\gamma_{\phi}(z)$ and observational constraints. For concreteness, we again use the Fang model as an example of Quintessence with variable $\lambda$. A wide, flat prior is chosen on $\tilde{\alpha}$, again to encompass the full range of EDE peaks. For K-essence we use a flat prior of $1 \leq  n \leq 2$. 

Rather than solve for the field variables, we solve for the perturbed fluid equations in \textsc{Camb},
\begin{eqnarray} \label{eqn:perts}
    \frac{d\rho_{\phi}}{d \tau} &=& -3 {\cal H} \gamma_{\phi}  \rho_{\phi}\,, \nonumber\\ 
    \frac{d \delta_{\phi}}{d\tau}&=&-\bigg[k u_{\phi}+\frac{\gamma_{\phi}}{2}\frac{d h}{d\tau}\bigg]-3 {\cal H} (c_\mathrm{s}^2-\gamma_{\phi}-1)\left( \delta_{\phi} + 3 {\cal H} \frac{u_{\phi}}{k} \right) - 3 {\cal H} \frac{1}{\gamma_{\phi} } \frac{d \gamma_{\phi}}{d\tau} \frac{ u_{\phi}}{k}  \,,\\ 
    \frac{d u_{\phi}}{d\tau}&=&-(1-3c_\mathrm{s}^2){\cal H}u_{\phi}+  \frac{1}{\gamma_{\phi} } \frac{d \gamma_{\phi}}{d\tau} u_{\phi}  + k c_\mathrm{s}^2\, \delta_{\phi}\,,
\end{eqnarray}
where conformal time $\tau$ is defined by $dt=a(\tau) d\tau$, ${\cal H} = a H$, $c_\mathrm{s}^2$ is defined in the rest-frame of the field, $\delta_\phi$ is the density perturbation, and $u_\phi \equiv \gamma_{\phi} v_\phi$ is the heat-flux, where $v_\phi$ is the velocity perturbation. The synchronous gauge metric perturbation is given by $h$. Note that the dynamics in terms of the field (and its perturbations) are equivalent to the perturbed fluid dynamics.

We perform a Markov Chain Monte Carlo (MCMC) analysis of the base \lcdm, Fang and K-essence models, using the public \textsc{Cobaya}~\citep{Torrado:2020dgo} and \textsc{Camb} codes~\citep{Lewis:1999bs}. We assume a spatially flat cosmology, with flat priors on the six base cosmological parameters, \\ $\left\{ H_0, \Omega_\mathrm{c} h^2, \Omega_\mathrm{b} h^2, n_\mathrm{s}, \log(10^{10} A_\mathrm{s}), \tau  \right\}$. The Fang model and K-essence models have an additional 4 parameters, $\left\{x_i, y_i, \tilde{\lambda}_i, \tilde{\alpha} \right\}$ and $\left\{x_i, y_i, \eta_i, n \right\}$ respectively. We use the ensemble sampler \textsc{emcee}~\citep{Foreman-Mackey:2012any} to sample over the model parameters, running 100 walkers in the ensemble. The minimum $\chi^2$ is then found by BOBYQA minimisation, using the chain best-fit as an initial guess. 

We use \Planck\ 2018 data~\cite{Planck:2018vyg} in combination with BAO data from BOSS DR12~\cite{BOSS:2016wmc}, 6dFGS~\cite{Beutler:2011hx} and SDSS-MGS~\cite{Ross:2014qpa}. The \Planck\ likelihoods used are the TT, TE and EE spectra at $\ell\ge 30$, the low-$\ell$ likelihood using the \textsc{Commander} component separation algorithm~\cite{Planck:2018yye}, the low$-\ell$ EE likelihood from the \textsc{SimAll} algorithm, and lensing~\citep{Planck:2018lbu}. We include the SH0ES $H_0$ measurement from ~\citep{Riess:2020fzl}\footnote{Note that the SH0ES measurement of $H_0 = 73.2 \pm 1.3\,{\rm km}~ {\rm s}^{-1}~{\rm Mpc}^{-1}$ is in the mid-range of local estimates. More recent SH0ES results~\cite{Riess:2021jrx} suggest a higher level of tension with~\lcdm, and even higher estimates exist from the Tully-Fisher relation (e.g.~\cite{Kourkchi:2020iyz}). Alternative local methods suggest a lower value, e.g.~\cite{Freedman:2020dne}. Since the purpose of this paper is to investigate possible resolutions to the $H_0$, we choose values in the mid-range.} and high $\ell$ CMB data from \textsc{Act} DR4~\citep{ACT:2020gnv}. Following the \textsc{Act} analysis, we exclude $\ell < 1800$ TT data to minimise double counting of information when combined with \Planck. \edit{In addition, we include the Pantheon SN sample~\cite{Pan-STARRS1:2017jku}.}

\begin{figure}
    \centering
    \includegraphics[scale=0.5]{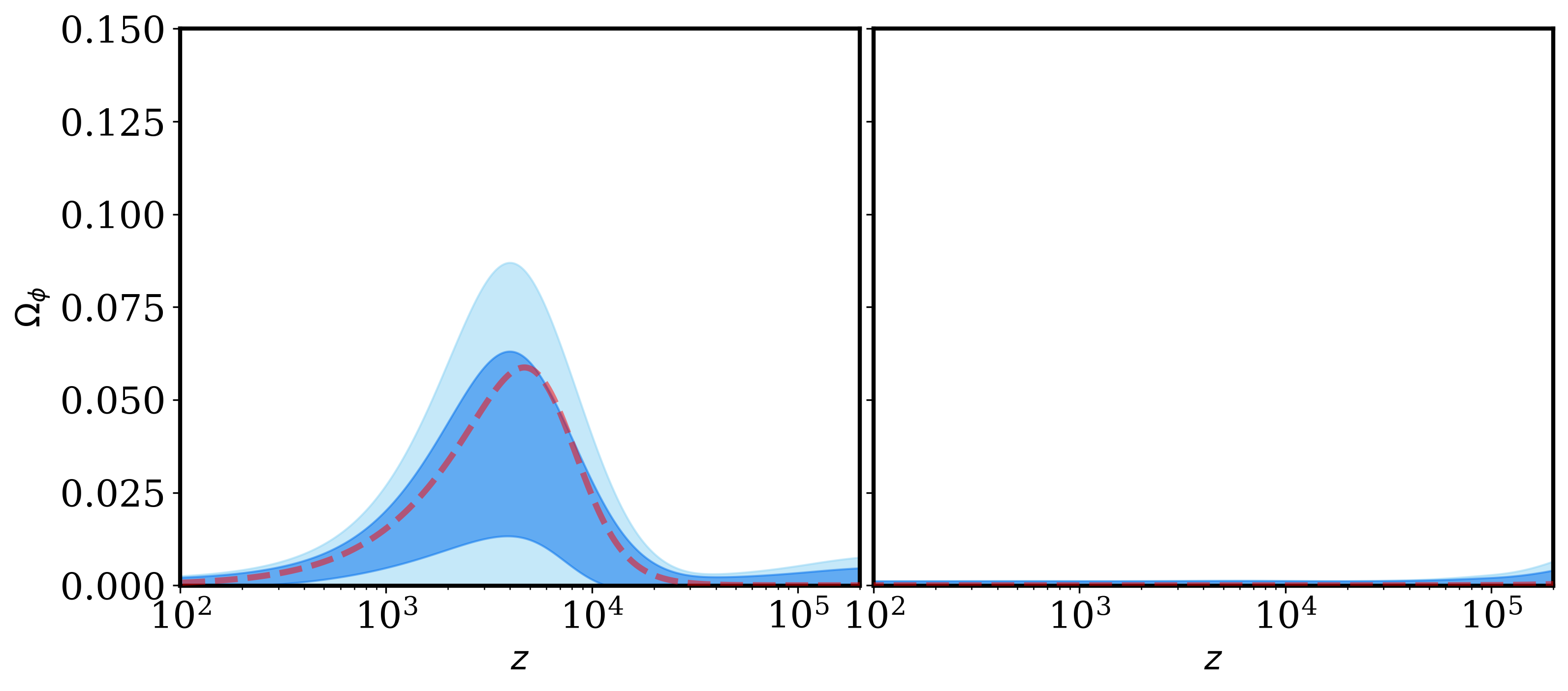}
    \caption{\footnotesize{Reconstruction of $\Omega_{\phi}(z)$ for the K-essence (left) and the Fang models (right). 1 and 2$\sigma$ confidences are indicated by the dark and light blue regions, and the best-fit by the dashed line. The best-fit of the K-essence model has $n=1.54$.}}
    \label{fig:recond_kessence_fang}
\end{figure}

The resulting parameter constraints on the base cosmological parameters and $\chi^2$ values are shown in Table.~\ref{table:parameters}. For the K-essence model, we find a $\Delta \chi^2=-14.2$ improvement over \lcdm, { competitive with other resolutions to the Hubble tension~\cite{Schoneberg:2021qvd}, but lower than the  axion model, which give $\Delta \chi^2 = -24.0$ for the same data combination~\cite{Poulin:2021bjr}. We attribute this to the differing background dynamics after the peak, and time/scale-dependent sound speed. } The main improvement of the K-essence model comes from the SH0ES measurement ($\Delta \chi_\mathrm{H0.riess2020}^2=-11$) and \textsc{Act} ($\Delta \chi_\mathrm{ACT}^2=-6$). Several features are consistent with previous work using the axion. First, whilst the best-fit $H_0=70.45\,{\rm km}~ {\rm s}^{-1}~{\rm Mpc}^{-1}$ is increased compared to \lcdm, it does not fully come into alignment with the SH0ES value. Second, there is an increase in the value of $S_8 \equiv \sigma_8 (\Omega_m/0.3)^{0.5}$,  where $\sigma_8$ is the matter clustering amplitude on scales of $8h^{-1}\,{\rm Mpc}$, which would increase tensions with the $S_8=0.790^{+0.018}_{-0.014}$ result from DES and KiDs~\cite{Kilo-DegreeSurvey:2023gfr}. The resulting reconstruction of $\Omega_{\phi}(z)$ is shown in Fig.~\ref{fig:recond_kessence_fang}. Similar to the axion, there is a preference for $\Omega_{\phi} \sim 0.06$ at $z \sim 4000$.  The posterior constraint on $n$ is $n=1.51^{+0.20}_{-0.28}$, giving a derived value of $c_s^2=0.53 \pm 0.13$. This is again similar to the axion model, which has $c_s^2 \approx 0.7$ over the relevant times and scales of interest~\cite{Smith:2019ihp}.

\def\arraystretch{1.2} 
\begin{table*}
\capstart
\centering
\resizebox{\textwidth}{!}{
\begin{tabular} {| l || c | c | c |} \hline

 Parameter & \lcdm &  K-essence &  Fang \\ \hline\hline

$H_0$ & $68.17\pm 0.37$ (68.09) & $69.7^{+1.2}_{-1.7}$ (70.45) & $68.19^{+0.38}_{-0.34}$ (68.29)\\
$\Omega_\mathrm{b} h^2$ & $0.02247\pm 0.00012$ (0.02247) & $0.02251^{+0.00014}_{-0.00018}$ (0.02251) & $0.02248\pm 0.00012$ (0.02250)\\
$\Omega_\mathrm{c} h^2$ & $0.11828\pm 0.00083$ (0.1185) & $0.1239^{+0.0046}_{-0.0062}$ (0.1278) & $0.11821\pm 0.00081$ (0.1180)\\
$n_\mathrm{s}$ & $0.9716\pm 0.0033$ (0.9711) & $0.9811\pm 0.0076$ (0.9873) & $0.9719\pm 0.0033$ (0.9722)\\
$\log(10^{10} A_\mathrm{s})$ & $3.056\pm 0.014$ (3.055) & $3.064\pm 0.015$ (3.058) & $3.055\pm 0.014$ (3.057)\\
$\tau_\mathrm{reio}$ & $0.0582\pm 0.0072$ (0.05784) & $0.0573\pm 0.0068$ (0.05122) & $0.0582\pm 0.0071$ (0.05884)\\ \hline
$r_d h$ & $100.50\pm 0.64$ (100.4) & $100.67^{+0.63}_{-0.75}$ (100.5) & $100.55\pm 0.63$ (100.7)\\
$S_8$ & $0.8177\pm 0.0097$ (0.8195) & $0.830\pm 0.014$ (0.8378) & $0.8168\pm 0.0092$ (0.8153)\\ \hline
$\chi^2_\mathrm{H0.riess2020}$ & 15.5 & 4.5 ({\bf -11.0}) & 14.2 ( -1.2)\\
$\chi^2_\mathrm{Planck}$ & 1014.2 & 1017.1 ( 2.8) & 1015.3 ( 1.1)\\
$\chi^2_\mathrm{ACT}$ & 240.4 & 234.4 ({\bf -6.0}) & 240.1 ( -0.3)\\ 
$\chi^2_\mathrm{BAO}$ & 5.2 & 5.2 ( -0.0) & 5.3 ( 0.0)\\
$\chi^2_\mathrm{SN}$ & 1034.8 & 1034.8 ( -0.0) & 1034.8 ( -0.1)\\ \hline
$\chi^2_\mathrm{data}$ & 2310.1 & 2296.0 ({\bf -14.2}) & 2309.7 ( -0.5)\\ \hline
Gaussian Tension ($\sigma$) & 4.1 & 3 & 4.0 \\ 
$Q_\mathrm{DMAP}$ Tension ($\sigma$) & 4.2 & 2.5 & 4.1 \\ \hline

\end{tabular}}
\caption[Mean (best-fit) parameter values for the \lcdm, K-essence and Fang models]{Mean (best-fit) parameter values for the \lcdm, K-essence and Fang models.}
\label{table:parameters}
\end{table*}

For the Fang model, $\Omega_{\phi}(z)$ is completely suppressed at all redshifts, with similar $\chi^2$ values to \lcdm. We note the model and parameter ranges are capable of producing peaks similar to K-essence, but in this case $c_s^2=1$. This underlies the importance of $c_s^2 < 1$ for a viable EDE model.

{
To assess the level of tension remaining with the SH0ES value and to facilitate the comparison with other models, we compute the Gaussian and $Q_\mathrm{DMAP}$ tension metrics. The Gaussian tension quantifies the difference in the posterior  to the SHOES measurement through

\begin{equation}
\frac{\bar{x}_\mathcal{D} - \bar{x}_\text{SHOES}}{(\sigma^2_\mathcal{D} + \sigma^2_\text{SHOES})^{1/2}}\,,
\end{equation}
where $\bar{x}_\mathcal{D}$ and $\sigma^2_\mathcal{D}$ are the posterior mean and standard deviation inferred from the dataset excluding SH0ES, and $\bar{x}_\text{SHOES}$ and $\sigma^2_\text{SHOES}$ are the SH0ES measurement. This is given in Table.~\ref{table:parameters}, where we observe a $4.1 \sigma$ tension in \lcdm\ and a $3 \sigma$ tension in the K-essence model. A downside of this metric is it may disfavor models with non-Gaussian posterior distributions, which is the case for scalar field models. A large volume of their parameter space results in no significant EDE, so marginalization over these shifts the posterior towards $\Lambda$CDM. This is illustrated in the reconstruction in Fig.~\ref{fig:recond_kessence_fang}, where the best-fit lies towards the upper end of the $1\sigma$ confidence region.

To circumvent this, we also compute the $Q_\mathrm{DMAP}$ tension~\cite{Raveri:2018wln}, which quantifies how the addition of the SH0ES measurement to $\mathcal{D}$ impacts the fit. It is given by
$Q_\mathrm{DMAP}=\sqrt{\chi^2_{\text{min},\mathcal{D}+SHOES} - \chi^2_{\text{min},\mathcal{D}}}$, and is equal to the Gaussian tension for Gaussian posteriors. The corresponding values are shown in Table.~\ref{table:parameters}. The two tension metrics are similar for $\Lambda$CDM and the Fang model, but $Q_\mathrm{DMAP}$ is reduced to $2.5\sigma$ for K-essence, which is again competitive with the models in ~\cite{Schoneberg:2021qvd}.
}